\documentclass[prl,twocolumn,superscriptaddress]{revtex4-2}
\usepackage{graphicx}
\usepackage{bm}
\usepackage{amssymb}
\usepackage{color}
\usepackage{amsmath}
\usepackage{mathrsfs}
\usepackage{amstext}
\usepackage[english]{babel}
\usepackage{latexsym}
\usepackage{array}
\usepackage{ulem}
\usepackage{multirow}         
\usepackage[usenames,dvipsnames]{xcolor}
\usepackage[colorlinks=true,citecolor=Blue,linkcolor=RubineRed,urlcolor=Blue]{hyperref}
\usepackage{physics}
\usepackage{enumerate}
\usepackage{array,etoolbox}
\preto\tabular{\setcounter{magicrownumbers}{0}}
\newcounter{magicrownumbers}

\newcommand{\expv}[1]{\langle #1 \rangle}
\newcommand{\pac}[1]{ \left\{ #1 \right\} }
\newcommand{\pap}[1]{\left( #1 \right)}
\newcommand{\pas}[1]{\left[#1 \right]}

\newcommand{\subl}[0]{{ \mathrm{C}}}
\newcommand{\supl}[0]{{ \mathrm{R}}}
\newcommand{\sub}[0]{{\rm SS}}

\newcommand{\Li}[0]{{\rm F}}
\newcommand{\sil}[0]{s_{\rm FC}}
\newcommand{\so}[0]{s_{\rm TP}}
\newcommand{\Hham}[0]{H_{\rm H}}
\usepackage{xspace}

\newtheorem{theorem}{Theorem}
\newtheorem{corollary}{Corollary}

\newtheorem{definition}{Definition}

\begin{document}

\title{Magnetization of electronic ground states in frustrated superstable graphs}
\author{F. P. M. M\'endez-C\'ordoba}
\email{fp.mendez10@uniandes.edu.co}
\affiliation{The Hamburg Centre for Ultrafast Imaging, Luruper Chaussee 149, Hamburg D-22761, Germany}
\affiliation{Universität Hamburg, Luruper Chaussee 149, Gebäude 69, D-22761 Hamburg, Germany}
\affiliation{Max Planck Institute for the Structure and Dynamics of Matter, Luruper Chaussee 149, 22761 Hamburg, Germany}
\affiliation{Departamento de F\'isica, Universidad de Los Andes, A.A. 4976, Bogot\'a, Colombia}
\author{J. Tindall}
\affiliation{Center for Computational Quantum Physics, Flatiron Institute, 162 5th Avenue, New York, NY 10010}
\author{D. Jaksch}
\affiliation{The Hamburg Centre for Ultrafast Imaging, Luruper Chaussee 149, Hamburg D-22761, Germany}
\affiliation{Universität Hamburg, Luruper Chaussee 149, Gebäude 69, D-22761 Hamburg, Germany}
\affiliation{Clarendon Laboratory, University of Oxford, Parks Road, Oxford OX1 3PU, UK}
\author{F. Schlawin}
\affiliation{The Hamburg Centre for Ultrafast Imaging, Luruper Chaussee 149, Hamburg D-22761, Germany}
\affiliation{Universität Hamburg, Luruper Chaussee 149, Gebäude 69, D-22761 Hamburg, Germany}
\affiliation{Max Planck Institute for the Structure and Dynamics of Matter, Luruper Chaussee 149, 22761 Hamburg, Germany}

\date{\today}
\begin{abstract}
Geometric frustration lies at the heart of many unconventional quantum phases in strongly interacting electron systems.
Here, we analytically determine the ground state magnetization of the half-filled Hubbard model on frustrated geometries where superstable states---eigenstates which are robust against frustration---are manifest. Our results apply to a broad class of lattices, including those where altermagnetic and superconducting states are known to emerge. Furthermore, they provide evidence for phase transitions involving a geometric rearrangement of magnetic correlations in the thermodynamic limit.
\end{abstract}
\maketitle

\textit{Introduction - } Magnetic frustration gives rise to many exotic phenomena in many-body physics, and is of fundamental importance in spintronics \cite{2002_Ritcher,2005_Ritcher,2008_Ritcher,FrustrMott_2018,FrustratedSpinSystems,Exp_Frust_review,Cold_Atoms_2023,Skyrmion_2021,Skyrmions_Spintronics,FrustratedSpintronics_2023,Victor_2024}. For instance, recent work has predicted flat band physics in a plethora of frustrated lattices \cite{FlatBands_2024} and materials with a Kagome lattice structure have shown evidence of intertwining between superconducting, density wave and spin liquid phases \cite{Kagome_Review}. In quasi-1D systems described by zigzag chains \cite{Zigzag_Caspers_1984,Zigzag_White_1996,Zigzag_exp_2025, dimerization_zigzag1D_2025}, frustration may be responsible for the spin-Peierls transitions witnessed in, e.g., $\mathrm{TTF\text{-}CuS}_4\mathrm{C}_4(\mathrm{CF}_3)_4 \text{ and } \mathrm{CuGeO}_3$~\cite{Spin-Peierls_TTF_1976, Spin-Peierls_CuGeO3_1993, Spin-Peierls_J1J2} whilst the Shastry-Sutherland lattice \cite{Shastry_Sutherland_Original,Shastry_Sutherland_Heisenberg,2024_Heisenberg_Altermagnet,Shastry_Sutherland_2025} is predicted to feature altermagnetism and could be realized, e.g., in $\alpha\text{- or }\kappa\text{-(BEDT-TTF)}_2\text{X}$ charge transfer salts \cite{Altermagnets_Hubbard_2024}. Charge transfer salts with other frustrated geometries also exhibit superconductivity, Mott physics, and putative spin liquid phases \cite{SC_ChargeTransfer_2024,PRB_Experimental_2024}. These results highlight the richness of frustrated quantum materials and motivate the development of numerical and analytical approaches to identifying the ground state properties of the corresponding many-body Hamiltonian.

In this regard, the Hubbard model and its exact~\cite{Uniform_Density_Lieb,Kubo_FiniteTemperature_1990,Tasaki_1992_SC_Corss,Oshikawa_LSM1D,Oshikawa_LSMAnyD,Tasaki_1998_Review,tasaki_Book} and numerical \cite{Hubbard_Computational,SC_Diamond,SpinLiquid_FrustratedHub_Numerical_2020, Hubbard2D_Frustrated_2025} treatments have played an outstanding role. Despite its apparent simplicity, the Hubbard model and extensions thereof qualitatively capture a wide range of many-body phenomena observed in highly distinct classes of quantum materials \cite{Kivelson_HubbardReview_2022, Kuzian_2023_Modeling}. These include Mott-insulating phases, high-temperature superconductivity \cite{Realistic_Params_ksalts,Kivelson_SC_Hubbard_2015,Review-Hubbard-Cuprates,SC_Hubbard_Plaquetes}, light-induced superconductivity \cite{Joey_Frank_Cavalleri_PhotoSC,Joey_Heating_n,Joey_PRB_SteadyStates,Joey_Quantum_Condensates,ProjectTheo,2021_Buzzi_PhaseDiagramLightInduced}, nontrivial topological properties \cite{Pollmann_Topo,Hubbard_Band_Topo} and magnetic order \cite{Ksalts_Triangular_2008_PRB,Ksalts_Triangular_2008,Ksalts_Triangular_2009,Ksalts_Triangular_2010,Magnetism_Hubbard}. Notably, exact results have revealed how magnetism in the ground state is intimately tied to the underlying graph that encodes the lattice geometry, hopping amplitudes, and electron filling \cite{Liebs_Theorem,Ferrimagnetic_Corrs,Tasaki_Nagaoka,Mielke_1991,Mielke_1999,Mielke_PRL,Mielke2012,2023_Kim_Nagaoka_Cactus,2024_Kim_Nagaoka}. In the half-filled case, these results apply only to bipartite graphs \cite{Liebs_Theorem,Ferrimagnetic_Corrs}. Specifically, Lieb's theorem \cite{Liebs_Theorem} guarantees uniqueness and determines the magnetization of the ground state in Hubbard models on such graphs, while the Shen-Qiu-Tian theorem \cite{Ferrimagnetic_Corrs} shows that this state exhibits ferrimagnetic correlations. Yet, as highlighted earlier, many important systems lack a bipartite lattice structure, giving rise to geometrical frustration \cite{IntroFrustration}. 

\begin{figure}[t!]
\begin{center}\includegraphics[width=1.0\linewidth]{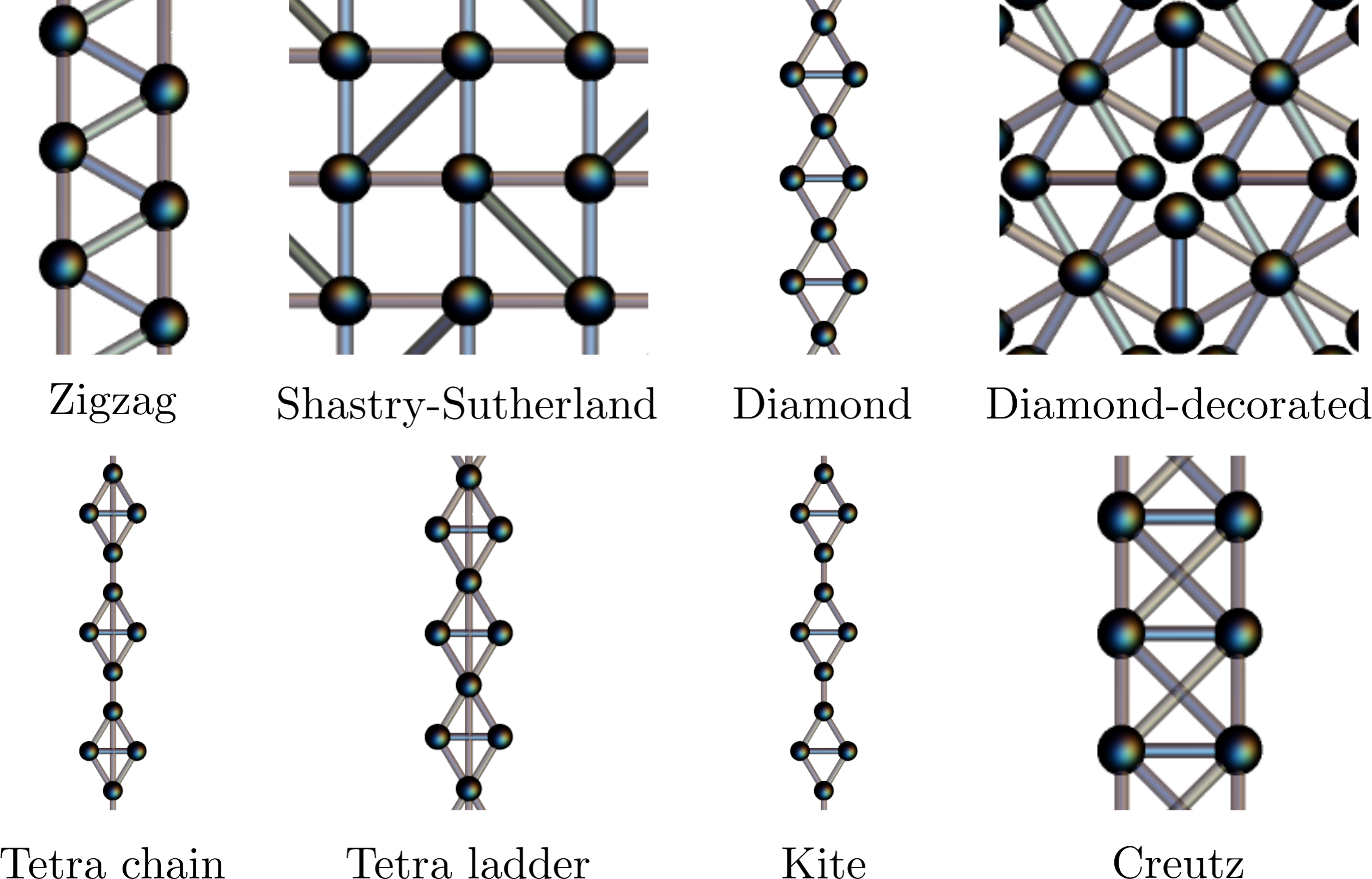}
\end{center}
\caption{Quasi-one-dimensional and two-dimensional examples of frustrated superstable graphs. 
Zigzag chains \cite{Spin-Peierls_J1J2}, as realized e.g. in $\mathrm{TTF\text{-}CuS}_4\mathrm{C}_4(\mathrm{CF}_3)_4 \text{ and } \mathrm{CuGeO}_3$. 
The Shastry-Sutherland lattice, expected to be realized in $\alpha\text{- or }\kappa\text{-(BEDT-TTF)}_2\text{X}$ \cite{Altermagnets_Hubbard_2024}. Examples of diamond-shaped compounds are $\mathrm{Cu_3(CO_3)_2(OH)_2}$, $\mathrm{K_3Cu_3AlO_2(SO_4)_4}$ and $\mathrm{Cu_3(CH_3COO)_4(OH)_2 \cdot 5H_2O}$ \cite{Azurite_2011_Magnetic, Diamond_2017_Magnetic, Diamond_2019_Magnetic}. Ref. \cite{FlatBands_2024} reports that, in total, more than two thousand known materials have diamond, tetra chain, tetra ladder, kite, or Creutz geometries. }
  \label{fig_exSS}
\end{figure}

In this paper, we extend Lieb's and Shen-Qiu-Tian's theorems within the strongly correlated regime of the half-filled Hubbard model to a broad class of non-bipartite graphs. In this limit of strong Coulomb repulsion, ground state properties may be deduced analytically from an associated spin Hamiltonian, which exhibits superstable ground states. These are eigenstates of the associated spin Hamiltonian, originally introduced by Shastry and Sutherland \cite{SS_Shastry_1983}, for which frustrating couplings exactly vanish. We refer to graphs that give rise to such superstable ground states in the associated spin Hamiltonian as superstable graphs. Fig. \ref{fig_exSS} shows notable examples of superstable graphs, including the zigzag and Shastry-Sutherland lattices, where the magnetic properties of the ground state of the Hubbard model are obtained via an effective Heisenberg Hamiltonian. In contrast, other superstable graphs---including the diamond chain \cite{Long_1990_TetramerDimerOpenNumerical,Takano_Triangular_Clusters,Triangular_Heisenberg} and the diamond-decorated square lattice~\cite{2020_Fukumoto_Diamond_Decorated,2023_DecoratedSquared_Jozef,2023_DecoratedSquared_Jozef_2,2023_DecoratedSquared_Jozef_3,Gerken_2025}---require incorporating additional charge fluctuations, as the Heisenberg model alone fails to reproduce the magnetic properties. 

As an application of our results, we study the Hubbard model on a diamond chain. This graph can be tuned from a bipartite to a frustrated regime where a superstable ground state emerges in the associated spin model. 
In the thermodynamic limit, this crossover develops into a first-order phase transition, a behavior that may reflect a more general characteristic of ground state phase diagrams of superstable graphs. We envision that our {\it analytical} results may guide the search for systems with topological spin textures or the study of superconducting condensation in driven quantum materials.

\begin{figure*}[t!]
\begin{center}\includegraphics[width=1.0\linewidth]{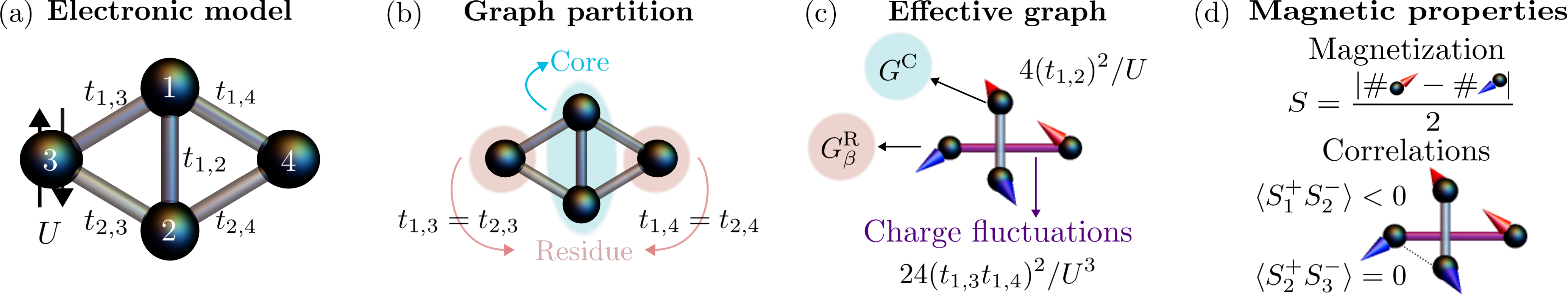}
\end{center}
\caption{
(a) Sketch of a Hubbard model on a single diamond with hopping weights $t_{i,j}:=t(e)$ with $e=(i,j)$ where $i$ and $j$ are vertices of the graph and on-site repulsion $U$. (b) The graph is partitioned into a core (formed by a dimer in this example) and a residue, identified through relations between the hopping weights. (c) The graph partition defines two bipartite graphs---$G^\subl$ and $G^\supl_\beta$---in an effective spin model, obtained in the strongly correlated regime. When the ground state of the effective model is superstable, couplings causing geometric frustration vanish. Consequently, the bipartition of $G^\subl$ and $G^\supl_\beta$ determines the ground state properties. The horizontal coupling connecting the vertices in $G^\supl_\beta$ is not present in the original graph; it is mediated by charge fluctuations arising from higher-order corrections, with weights parametrized by a hopping-anisotropy $\beta$. (d) The ground state magnetization, $S$, and the sign of the magnetic correlations, $\expv{S^+_iS^-_j}$, of the Hubbard model are given by the partition of the effective graph.}
  \label{fig_graphs}
\end{figure*}
\textit{Results -}
We study magnetism in Hubbard models on weighted graphs [see Fig. \ref{fig_graphs}(a)].  
A weighted graph $G = \pap{V, E, t(E)}$ is a set of vertices $V$, a set of edges $E$ which correspond to pairs of vertices $(i,j)$ with $i \neq j$ and $i, j \in V$, and weights $t(e)$ which maps edges $e\in E$ into nonzero scalars. For the Hubbard model, the weights represent the hopping amplitude between pairs of sites. The Hubbard model on a weighted graph thus reads
\begin{equation}
\label{eq:HubbardGeneralLattice}
    H(G)=U\sum_{i\in V}n_{i,\uparrow}n_{i,\downarrow}-\sum_{e=\pap{i,j} \in E}\sum_{\sigma}t(e) \left(c_{i,\sigma}^\dagger c_{j,\sigma} + {\rm h.c.} \right),
\end{equation}
where $n_{i,\uparrow}$, $c_{i,\sigma}$ ${c_{i,\sigma}^{\dagger}}$ are the fermionic number, annihilation, and creation operators at vertex $i\in V$ with spin $\sigma \in \pac{\uparrow,\downarrow}$. The interaction $U>0$ accounts for the local Coulomb repulsion. The Hamiltonian has a SU(2) global spin rotation symmetry as it commutes with all the total spin operators $ S^\alpha=\sum_{i\in V} S^\alpha_i$ where $S^\alpha_i=\sum_{\sigma,\sigma^\prime}c_{i,\sigma}^{\dagger}(\sigma^\alpha)_{\sigma \sigma^\prime}c_{i,\sigma^\prime}$ with $\sigma^\alpha$ the Pauli matrices and $\alpha \in \{x,y,z\}$ \cite{Essler_1dHubbard_2005}. This symmetry allows the characterization of every eigenstate of the Hubbard Hamiltonian with a total spin $S$, which can take positive integer or half-integer values. 

At half-filling, in the strongly correlated regime  [i.e., $U\gg\max(|t(e)|)$, see End Matter (EM)], the ground state magnetic properties of Eq. \eqref{eq:HubbardGeneralLattice} can be obtained  from effective spin models derived via a Schrieffer-Wolff (SW) transformation expansion up to order $n$ in $U^{-1}$ \cite{t-U_Expansion,supplementary_material}.  The $n=1$ order  gives $\Hham\pap{G_{\rm H}}$; the spin-1/2 Heisenberg Hamiltonian on the weighted graph $G_{\rm H}:=\pap{V, E, 4t(E)^2/U}$ (i.e., the parent graph with transformed weights) and $\Hham\pap{G}:=\sum_{e=\pap{i,j} \in E}t(e)(\Vec{S}_i \cdot \Vec{S}_j-1/4)$. This effective model is known to capture the magnetic properties of Hubbard models on connected bipartite graphs \cite{defconn, tasaki_Book}. A graph is bipartite if we can define disjoint sets $A$ and $B$ such that $A \sqcup B = V$ and for every edge $e = (i,j) \in E$, we have $i \in A$ and $j \in B$. According to Lieb's, Lieb-Mattis' and  Shen-Qiu-Tian's theorems, this partition of vertices determines the total spin of both models as $S=(||A|-|B||)/2$ \cite{Lieb_Theorem_Heisenberg,Liebs_Theorem} and the sign of their ground state magnetic correlations \cite{Ferrimagnetic_Corrs,Tian1994CoexistenceFerroAntiFerro}, $\expv{S^+_iS^-_j}$, where $S^\pm_i=S^x_i\pm iS^y_i$. In other words, these theorems map the problem of finding the ground state magnetic properties to a simple graph theory problem---finding the bipartition.

In contrast to the bipartite case, higher orders of the SW transformation expansion may be necessary to capture the magnetic properties of the Hubbard model on frustrated graphs. At these higher orders, the resulting spin models do not `live' on the same graph as the parent Hubbard model: they generically feature new pairwise interactions that add edges to the parent graphs, altering their geometry, and they also include beyond-pairwise interactions that cannot be represented as edges \cite{BeyondPairwiseInteractions}. However, on superstable graphs, those frustrating and beyond-pairwise interactions vanish (see EM), and the ground state magnetic properties can be found from effective spin models on bipartite graphs. Hence---in the same spirit as the previous theorems---determining magnetic properties on superstable graphs reduces to finding effective bipartite graphs.


To see how this occurs, consider the $n=1$ order of the SW transformation expansion of the superstable graph in Fig.~\ref{fig_graphs} (a): a diamond composed of a vertical dimer uniformly coupled ($t_{1,3}=t_{2,3}$ and $t_{1,4}=t_{2,4}$) to residual vertices ($3$ and $4$) that introduce frustration [see Fig.~\ref{fig_graphs} (b)]. Due to the uniform coupling, the frustrating spin interactions are proportional to the total spin of the dimer, $S_1^\alpha+S_2^\alpha$, and thus vanish when it forms a singlet, $\ket{0}_{\subl}:=(\ket{\uparrow}_1\ket{\downarrow}_2-\ket{\downarrow}_1\ket{\uparrow}_2)/\sqrt{2}$. Consequently, the states $\ket{0}_{\subl}\otimes\ket{\sigma}_3\otimes\ket{\sigma^\prime}_4$ are said to be superstable to the frustrating terms \cite{SS_Shastry_1983}. When $t_{1,2}$ is sufficiently large, these become ground states with $S=0$ and $1$, corresponding to a singlet or a triplet on the residual vertices, respectively. 
However, charge fluctuations from the $n=3$ order of the SW transformation expansion break this degeneracy in the Hubbard model as they mediate an antiferromagnetic coupling $\sim  t_{1,3}^2 t_{1,4}^2 / U^3$ 
between the residual vertices, thereby forming an effective dimer on sites $3$ and $4$, and energetically favor its singlet state [see Fig. \ref{fig_graphs} (c)]. Hence, the $n=3$ corrections stabilize a unique superstable ground state with $S=0$. 
Due to the superstable property of the ground state, the magnetization of this state can be thought of as the sum of the ground state magnetization of two independent dimers. We thus see that the ground state magnetic properties of the frustrated graph in Fig. \ref{fig_graphs}~(a) are found from the analysis of a Heisenberg Hamiltonian on a different bipartite graph, as shown in Fig. \ref{fig_graphs}~(d). 


This simple approach generalizes to a broader class of superstable graphs. Importantly, in the EM, we introduce an algorithm based on established graph theory concepts, such as modular decomposition~\cite{ModularDecomposition,2010_ModDecAlgorithms}, that systematically identifies graphs that show superstable ground states in the $n=1$ or $n=3$ order.
Given a superstable graph, the algorithm returns two effective bipartite graphs: a core graph $G^\subl$---generalizing the first dimer---whose connected components \cite{conn_comp_def} form connected bipartite and balanced (i.e., $|A|=|B|$) graphs, and a residue graph $G_\beta^\supl$---built from the residual vertices and the effective edges created by charge fluctuations---which contains at most one unbalanced connected component. In the single-diamond example above, $G_\beta^\supl$ is the dimer on sites $3$ and $4$. The weights of the residue graph depend on the hopping-anisotropy $\beta:=\{\max_{e\in E \setminus E^\subl} [|t(e)|]/\min_{e\in E^\subl} [|t(e)|]\}^2$, with $E^\subl$ the set of edges of $G^\subl$. Physically, the hopping-anisotropy controls how strongly the residue frustrates the core singlets. In the single-diamond example above, $\beta=\{\max(|t_{1,3}|,|t_{1,4}|)/|t_{1,2}|\}^2$, so small $\beta$ corresponds to a strongly bound vertical dimer, stabilizing its singlet and effectively decoupling it from the residual vertices. We emphasize that this residue graph contains the new edges added by charge fluctuations [see Fig. \ref{fig_exSS} (c)], which depend on the geometry of the parent graph (see EM). The bipartition of these graphs determines the ground state magnetic properties, as given by our main results: 

\begin{theorem}
If a graph is superstable, then, in the strongly correlated regime, for sufficiently small $\beta$, every ground state has total spin $S=(||A|-|B||)/2$. The ground states are exactly 
$(2S + 1)$-fold degenerate. Here, $ A\sqcup B$ is the bipartition for the only unbalanced component of $G^\supl_\beta$. If all are balanced, $S = 0$. 
 \end{theorem}
Here, as above, the strongly correlated regime means that $U$ is large enough so that the gap of a unique ground state (up to the trivial degeneracy) of the order-$n$ SW effective spin Hamiltonian is not closed by higher-order corrections. Thus, the total $S$ predicted by this ground state is the same as that of the Hubbard model. This regime always exists; see the EM for the formal statements.
The trivial $2S+1$ degeneracy comes from the SU(2) invariance of the model, which guarantees that each eigenenergy is at least $(2S+1)$-fold degenerate. Therefore, the ground states are unique apart from this mandatory spin degeneracy. Note that the total magnetization is given by the sole unbalanced component of the residue graph. As all other components of the effective graphs are balanced, they cannot contribute to the total magnetization.

 Under the same conditions of Theorem 1,
 ferrimagnetic correlations \cite{tasaki_Book} appear within the individual components of the effective graphs. Formally, this statement is expressed by the following corollary.
\begin{corollary}[Approximate ferrimagnetism]
Within the bipartition of
each component of $G^\subl$ and $G^\supl_\beta$, up to $1/U^2$ corrections, the magnetic correlations, $\expv{S^+_iS^-_j}$, are positive (negative) when $i$ belongs to the same (opposite) disjoint partition as $j$ and zero when $i$ and $j$ belong to different components.
\end{corollary}
The proofs of the theorem and corollary are in the EM.

{\it Applications - } Although the emergence of a superstable ground state in the effective spin model might seem uncommon at first glance, important lattice structures host this property. For example, bipartite graphs are naturally superstable (there is no frustration), and Theorem 1 reduces to Lieb's theorem. In this case, Theorem 1 applies to any $U>0$ \cite{Liebs_Theorem}, the parameter $\beta$ can have any positive value, the graph $G^\subl$ is empty, and $G^\supl_\beta=G_{\rm H}$. Under the same conditions, Corollary 1 reduces to the Shen-Qiu-Tian theorem \cite{Ferrimagnetic_Corrs}, where the corrections are exactly zero. Thus, our results correctly reproduce the ferrimagnetism of bipartite graphs through the $n=1$ expansion. Meanwhile, the properties of the zigzag chain \cite{Zigzag_Caspers_1984,Zigzag_White_1996,Zigzag_exp_2025} and the Shastry-Sutherland lattice \cite{Altermagnets_Hubbard_2024} are also described by this expansion order, representing the first set of graphs where our analytical results go beyond Lieb's and Shen-Qiu-Tian theorems. For these lattices, all vertices are part of a diagonal dimer, which together form $G^\subl$. Instead, the graph $G^\supl_\beta$ is empty, and Theorem 1 yields $S=0$. This result, together with Corollary 1, predicts that the dimerization observed in Heisenberg models \cite{Shastry_Sutherland_Heisenberg,Zigzag_White_1996} on these geometries also occurs in the Hubbard model. 

\begin{figure*}
    \centering
    \includegraphics[width=1\linewidth]{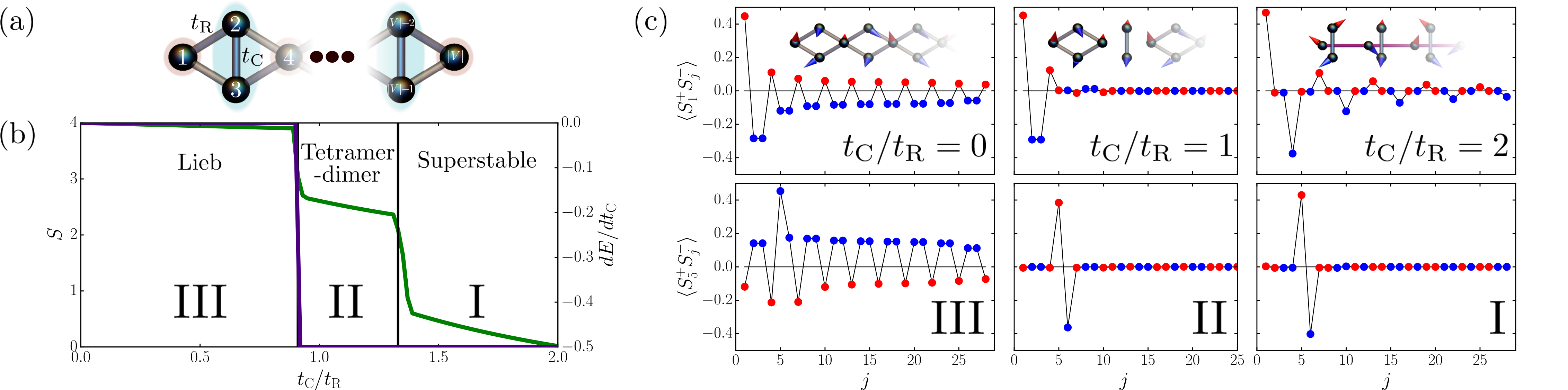}
    \caption{(a) Diamond chain of $\vert V \vert$ vertices and its partitioning into cores (vertical shading) and residue (circles). The weights of the vertical (diagonal) hoppings are given by $t_{\subl}$ ($t_\supl$). (b) Ground state total spin $S$ (purple line) and first derivative of the energy $dE/dt_{\rm C}$ (green line) of the Hubbard model on the diamond chain as a function of the strength of the hopping ratio $t_{\subl}/t_\supl$. The vertical lines mark changes in the ground state characteristics, separating three ground state regimes. ($\mathrm{I}$) The superstable regime  is directly described by our results. ($\mathrm{II}$)-($\mathrm{III}$) The tetramer-dimer and Lieb regimes are related to our results through adiabatic continuation. (c) Representative correlations in the three regimes. The insets show the effective bipartite graph that describes the magnetization and correlations in each region. The parameters are $|V|=28$ and $U=6t_\supl$.
 } 
\label{fig_Triang}
\end{figure*} 

Furthermore, several candidate flat-band materials, where Hubbard physics is expected to dominate, may also host superstable ground states, provided that the hopping-anisotropy is sufficiently small, as their lattice structures are described by superstable graphs. Ref. \cite{FlatBands_2024} identifies, inter alia, the twelve most prevalent frustrated flat-band lattices among more than one hundred thousand materials from the Materials Project database \cite{Materials_Project}. We have identified that five of these non-bipartite structures (i.e., the diamond, tetra chain, tetra ladder, kite, and the Creutz ladder graphs in Fig. \ref{fig_exSS}) are superstable graphs, corresponding to more than two thousand known materials with the potential to host superstable ground states.

In the following, we explore the Hubbard model on one of these common lattices, the diamond chain, where we need to resort to $n=3$ to appropriately describe the magnetic properties.
Our results explain density matrix renormalization group (DMRG) simulations of the ground state of the model~\cite{SCHOLLWOCK,Or_s_2019DMRG}.
The vertex labeling is shown in Fig.~\ref{fig_Triang} (a). For simplicity, we assume that all the vertical edges have weights $t_\subl$ and the diagonal edges have nonzero weights $t_\supl$. For nonzero $t_\subl$, the hopping-anisotropy is $\beta=(t_\supl/t_\subl)^2$, each pair of vertices coupled with a vertical hopping constitutes a core, and the centered sites are the residual vertices. Fig.~\ref{fig_Triang} (b) shows the total spin and different ground state regimes of the model obtained from converged DMRG calculations. The crossover between distinct regimes is identified in our simulations by discontinuities in the first derivative of the energy with respect to the ratio $t_{\subl}/t_\supl$. The insets illustrate typical magnetic correlations within each regime. 

 Regime $\mathrm{I}$, occurring at $t_\subl / t_\supl \gtrsim 1.34$, is the superstable regime, where Theorem 1 and Corollary 1 determine the magnetic properties. The sketch of the effective bipartite graph for this regime is shown in Fig. \ref{fig_Triang}~(c). From the sketch, we see that the diagonal edges of the parent graph, which frustrate the vertical dimers, are removed by the superstable ground state. Consequently, this creates disconnected vertical dimers in the effective graph, which together form the graph $G^\subl$. 
The nonzero horizontal weights in the effective spin model are added by the $n=3$ order of the SW transformation expansion. They create a balanced one-dimensional chain composed of the residual vertices ($G^\supl_\beta$). 
Since all the connected components of the effective graph are balanced, the total spin is $0$ according to Theorem 1. The sign of the numerically obtained correlations agrees with the predictions of Corollary 1 applied to the sketched graph: the one-dimensional chain is disconnected from the dimers (and vice versa), and the correlations are antiferromagnetic within the connected components of the obtained graph. 
Thus, these results illustrate that the magnetic correlations behave as a composite of ferrimagnetic contributions from the connected components.
Note that the expansion to $n=3$ is crucial here: At $n=1$, the SW transformation gives the Heisenberg model on a diamond chain (with the weights determined by $G_{\rm H}$), which is known to show superstable ground states for sufficiently small $\beta$  \cite{SS_Shastry_1983,Takano_Triangular_Clusters,Triangular_Heisenberg}. These states are described by the ground state of the Heisenberg model on a bipartite graph with vertical dimers, but the residual vertices are completely isolated from any other vertex, as in our example of the single diamond.
Consequently, $n=1$ gives highly degenerate (proportional to the system size) ground states, which would result in a paramagnetic response of the system to an external magnetic field, whereas $n=3$ correctly predicts antiferromagnetism. The latter shows explicitly how the Heisenberg model predicts incorrect physics for frustrated superstable graphs, as it ignores charge fluctuations that lift the ground state degeneracy. 

The other two regimes are understood via adiabatic continuation \cite{Adiabatic_Theorem} between the ground state of $H(G)$ and the superstable ground state of the Hubbard model on another superstable graph, $G^\prime$ \cite{supplementary_material}. The graph $G^\prime$ that enables the adiabatic continuation in regime $\mathrm{II}$ is obtained by removing every second vertical edge \cite{supplementary_material}. The resulting superstable ground state eliminates diagonal edges connected to the vertical dimers, leading to a balanced effective graph composed of intercalated tetramers (diamond-shaped bipartite formations of four vertices) and dimers, as shown in the sketch for the corresponding regime in Fig. \ref{fig_Triang} (c). Analogously, for regime $\mathrm{III}$, the superstable graph $G^\prime$ is bipartite and obtained by removing all vertical edges (i.e., setting $t_\subl=0$)~\cite{supplementary_material}. Remarkably, we observe that Theorem 1 and Corollary 1, applied to the effective graphs obtained through adiabatic continuation, correctly predict the magnetization and the sign of the numerically obtained correlations, although these results only apply directly to regime $\mathrm{I}$.

The three regimes were shown to converge into ground state phases separated by first-order quantum phase transitions in the thermodynamic limit in Ref. \cite{ProjectTheo}. The emergence of such first-order quantum phase transitions is rooted in Lieb's theorem and Theorem 1. At $t_\subl=0$ the graph is connected bipartite, and Lieb's theorem determines $S\propto L/2$ ($L$ is the number of unit cells) as each unit cell is unbalanced due to a single vertex. Instead, the ground state magnetization is $S=0$ in the superstable regime (assuming an even number of unit cells) as the effective spin model describes a balanced one-dimensional chain with disconnected balanced dimers. Since the ground states belong to two different symmetry sectors, these results imply the existence of at least one ground state energy level crossing at a given system size as $t_{\subl}/t_\supl$ increases. 
As shown for the diamond chain, there are two energy level crossings from Lieb's regime to the superstable regime. These crossings prevail when increasing the system size and lead to first-order quantum phase transitions.

Beyond the diamond chain, such a level crossing will emerge whenever a hopping-anisotropy ratio analogous to $t_{\subl}/t_\supl$   interpolates between a bipartite and a superstable regime. Specifically, provided that the total spin predicted by Lieb's theorem in the bipartite regime and Theorem 1 in the superstable regime differ, for sufficiently large $U$, there is at least one ground state energy level crossing as a function of  the hopping-anisotropy ratio. This energy level crossing is expected to give rise to a quantum phase transition \cite{Sachdev_QPT,FirstOrder_PRL} , which need not be first-order, provided that a notion of increasing system size can be defined on the superstable graph, and that the superstable regime remains finite and converges in the thermodynamic limit. 

 
{\it Conclusions - }
Our results offer new analytical insights into the magnetic properties of frustrated correlated itinerant electron systems and the emergence of quantum phase transitions in the thermodynamic limit. Specifically, we have demonstrated the existence of a ground state regime where the ground state magnetization and ferrimagnetic correlations of a relevant class of frustrated lattices (i.e., with example applications in altermagnetism, flat-band physics, etc.) are determined from a bipartite graph. In this regime, charge fluctuations can become crucial in shaping the magnetic properties as they lift degeneracies. This mechanism has been shown to stabilize topological spin textures in itinerant magnetic systems \cite{Hayami_2021_ItinerantMagnets}, suggesting that similar topological features may be identified in frustrated geometries with our results or even induced through photoinduced frustration modulation~\cite{2025_Mitrano_PhotoinducedFrustration}. 

Our findings are further important for the study of driven quantum materials more broadly~\cite{SentefRMP}, since ferrimagnetic ground states were recently identified as promising candidates for optically inducing superconducting condensation in a maximum entropy state upon driving~\cite{Tindall2021liebstheorem}.
In this context, superstable states allow a straightforward interpretation of the simulations in \cite{ProjectTheo}, as well as the observations of transient superconducting signals in \cite{Joey_Frank_Cavalleri_PhotoSC, 2021_Buzzi_PhaseDiagramLightInduced}. 
Hence, our results can guide both the development of a symmetry-based theoretical understanding of these phenomena and the search for new opportunities for optical control.




\section{Acknowledgments}
We thank Bill Sutherland and David DiVincenzo for useful communications. We thank Elliott H. Lieb for reading our manuscript. DMRG calculations were performed with the TeNPy~\cite{tenpy} library in the University of Oxford Advanced Research Computing (ARC) facility. FPMMC,
DJ and FS acknowledge support from the Cluster of Excellence `Advanced Imaging of Matter' of the Deutsche Forschungsgemeinschaft (DFG) - EXC 2056 - project ID 390715994. JT is grateful for ongoing support through the Flatiron Institute, a division of the Simons Foundation.
\newpage

\section*{End Matter}
\subsection{The strongly correlated regime}
\begin{definition}[Strongly correlated regime]
The strongly correlated regime is where $U$ is sufficiently large so that the quantum number $S$ can be continued to the $S$ of a unique ground state predicted by the order $n$ expansion in $t/U$ of a Schrieffer-Wolff transformation. 
\end{definition}
This regime exists since, for sufficiently large $U$, the  SW transformation expansion series is absolutely convergent (see Refs. \cite{supplementary_material,BRAVYI_SWT}) and the ground state predicted from the order $n$ expansion is robust to higher order terms \cite{BRAVYI_SWT,Kato1995,tasaki_Book}.
\subsection{Effective spin Hamiltonian of a superstable graph}
A superstable graph yields, at some order $n$ in the SW transformation expansion and up to a scale factor, an effective spin Hamiltonian of the form
\begin{equation}
\label{Eq_GEN_SSProblem_NOSUBG}
H^\sub(\beta)=\Hham\pap{G^\subl}+ \Hham\pap{G^\supl_\beta} 
+\sum_{i=1}^{(n+1)/2}\sum_{q=1}^{Q_i}\beta^{i} \mathcal{X}_{i,q} \mathcal{Y}_{i,q},
\end{equation}
where the graphs $G^\subl$ and $G^\supl_\beta$ must be unions of connected bipartite graphs. Each connected component of $G^\subl$ must be balanced, while at most one of the connected components of $G^\supl_\beta$ can be unbalanced. Although $G^\supl_\beta$ depends on $\beta>0$, this parameter only rescales the weights without changing the vertex or the edge sets. The third term encodes the frustration in the system. The $\mathcal{X}_{i,q}$ ($\mathcal{Y}_{i,q}$) operators act on the subspace defined by the vertices of $G^\subl$ ($G^\supl_\beta$). The total number of interactions for a given order $i$ is $Q_i$, which depends on the edges connecting the components of $G^\subl$ and $G^\supl_\beta$. To ensure the frustrating terms vanish, we require that the ground state of $\Hham\pap{G^\subl}$ is superstable to all $\mathcal{X}_{i,q}$. Formally, this means the ground state of $\Hham\pap{G^\subl}$ is an eigenstate of the operators $\mathcal{X}_{i,q}$ with eigenvalue 0~\cite{SS_Shastry_1983}. Consequently, the latter property enforces that the ground states of $\Hham(G^\subl)+ \Hham(G^\supl_\beta)$ are superstable to the frustrating term in Eq. \eqref{Eq_GEN_SSProblem_NOSUBG}.


\subsection{Algorithm to identify superstable graphs}
\label{Sec-SSNBGRAPHSExact}
We describe a constructive algorithm that, given an input graph $G$ [cf. Fig. \ref{fig_graphs} (a)], identifies features that lead to a Hamiltonian of the form of Eq. \eqref{Eq_GEN_SSProblem_NOSUBG}. It relies on the identification of substructures of the graph, which we call cores. These structures are characterized through induced subgraphs \cite{diestel2025graph_Theory}: we denote an induced subgraph of $G$ as $G\pas{M}=(M,E^M,t(E^M))$, where $G\pas{M}$ takes all the edges and weights of $G$ where both ends are elements of $M\subseteq V$. With this, we formally define 
\begin{definition}[Core]
A subset $X\subseteq V$ is a core if the following conditions hold: 1. The induced subgraph $G\pas{X}$ is connected and balanced. 2. One of the following holds for each $v \in V\setminus X$: {\bf (i)} $v$ is not part of another core, and the weights from $v$ to all vertices in $X$ are equal, denoted by $t_X(v)$. That is, $t(v,x)=t_X(v)\forall x\in X$. {\bf (ii)} $v$ belongs to another core $X^\prime$, and the weights connecting $X$ and $X^\prime$  satisfy the relation $\pas{t(x,v)}^2=\pas{t_{X-X^\prime}(x)}^2+\pas{t_{X^\prime-X}(v)}^2$ $\forall x\in X$ and $\forall v\in X^\prime$. Here, $t_{X-X^\prime}(x)$ represents a fixed weight contribution that depends on the module pair $(X,X^\prime)$ but not on the specific vertex $v\in X^\prime$. Similarly, $t_{X^\prime-X}(v)$ represents the contribution that depends on $v$, but is independent of the specific $x\in X$.
\end{definition}
The function $t_{X-X^\prime}(x)$ captures how the weights connecting $X$ to $X^\prime$ depend on the choice of vertex in $X$, but remain independent of the specific vertex in $X^\prime$. This structure ensures that part of the weights connecting the two cores are uniform across all vertices in $X^\prime$, guaranteeing superstability \cite{supplementary_material}. Similarly, the function $\pas{t_{X^\prime-X}(v)}$ ensures uniform connections between each vertex in $X^\prime$ and all the vertices in $X$. With the definition, we describe the algorithm below.

First, identify the presence of cores in the input graph and denote them as sets $\subl_u$, where $u\in \mathcal{I_\subl}$ with $\mathcal{I_\subl}$ the index set labeling the cores. This can be achieved through a generalized modular decomposition algorithm~\cite{ModularDecomposition, 2010_ModDecAlgorithms} that incorporates conditions 1 and 2.~(ii) in the definition of a module.
After identifying the cores, define the residue $V^\supl:=V\setminus \sqcup_{u\in \mathcal{I_\subl}}\subl_u$ [see Fig. \ref{fig_graphs} (b)]. This yields the partition: 
\begin{equation}
\label{Eq:Prop1Lattice}
    G=\pap{\bigoplus_{u\in \mathcal{I_\subl}} G\pas{\subl_u}}\oplus G\pas{V^\supl}\oplus(\varnothing,E^\Li,t(E^\Li)),
\end{equation} 
where $\oplus$ denotes the disjoint union of graph elements, and the set of edges $E^\Li:=E\setminus[(\sqcup_{u\in \mathcal{I_\subl}}E^{\subl_u})\sqcup E^\supl]$ are those generating frustration between cores or cores and residue. 

Then verify whether $G\pas{V^\supl}$ is bipartite with at most one unbalanced connected component. In that case, $G$ is superstable and  $n=1$ is the order of the SW transformation expansion that leads to Eq. \eqref{Eq_GEN_SSProblem_NOSUBG} \cite{supplementary_material}. Then, $\bigoplus_{u\in \mathcal{I_\subl}} G\pas{\subl_u}\equiv_{t(e)\neq 0} G^\subl$ and $G\pas{V^\supl}\equiv_{t(e)\neq 0} G^\supl_\beta$, where $\equiv_{t(e)\neq 0}$  denotes equality up to weight magnitude: the vertex and edge sets coincide, but the actual weight values may differ.

If $G\pas{V^\supl}$ is only bipartite, higher orders in the expansion may lead to Eq. \eqref{Eq_GEN_SSProblem_NOSUBG}. In that case, check the following additional conditions that yield Eq. \eqref{Eq_GEN_SSProblem_NOSUBG} for the $n=3$ expansion order \cite{supplementary_material}; (i) none of the connected components of $G\pas{V^\supl}$ or $G\pas{\subl_u}$ can have more than two vertices; (ii) the weights coupling different cores have to follow the additional constraint $t(x,v)=t_X(v)=t_{X^\prime}(x)=t_{X-X^\prime}=t_{X^\prime-X}$ for $x\in X$, $v\in X^\prime$, with $X \neq X^\prime$; (iii) the $n=3$ SW transformation expansion adds edges with ends in $V^\supl$ that turn the effective spin model version of the subgraph $G\pas{V^\supl}$ into $G^\supl_\beta$, which is a bipartite graph with at most one unbalanced component. These added edges depend on the walks (routes following adjacent vertices) of the graph $G$ involving edges in $E^\Li$. Only simple two-paths and four-cycles contribute to the $n=3$ expansion order. Let $i,j,k,l \in V$ be distinct vertices. A simple two-path consists of edges $\pap{i,j}$ and $\pap{j,k}$. A four-cycle is a set of four edges $\pap{i,j},\pap{j,k},\pap{i,l},\pap{l,k}$. The $n=3$ expansion adds an edge between a pair of $V^\supl$ vertices  for each two-path or four-cycle that is formed exclusively from edges in $E^\Li$ [cf. purple edge in Fig. \ref{fig_graphs}~(c)]. If, when added to $G\pas{V^\supl}$, these new edges generate a bipartite graph, the input graph is superstable and we can determine the magnetic properties from the bipartition of the different $G\pas{\subl_u}$, and $G\pas{V^\supl}$ with the added edges [cf. Fig. \ref{fig_graphs}~(d)].



\subsection{Proof of Theorem 1}
\label{Sec_ProofTheorem}
Consider the Hamiltonian in Eq. \eqref{Eq_GEN_SSProblem_NOSUBG} and let $G^\subl=\bigoplus_{u\in \mathcal{I}}  G^\subl\pas{\subl_u}$ and $G^\supl_\beta=\bigoplus_{v\in \mathcal{I}}  G^\supl_\beta\pas{\supl_v}$. The set $\mathcal{I}$ is the index set labeling the components of the corresponding subgraphs. 
The $\subl_u\subseteq V^\subl$ ($\supl_v\subseteq V^\supl$) are such that $V^\subl=\bigsqcup_{u\in \mathcal{I}}\subl_u$ ($V^\supl=\bigsqcup_{v\in \mathcal{I}}\supl_v$). 
We construct a basis given by states $\ket{j,k,\beta}:=\ket{j}_\subl \otimes \ket{k,\beta}_\supl$. Here, $\ket{j}_\subl$ are all the eigenstates of $\Hham\pap{G^\subl}$ with eigenvalue $\lambda^\subl_j$ with $j\in \pas{0,2^{|V|^\subl}-1}\subset \mathbb N$ ordered with increasing energy. 
According to the Lieb-Mattis theorem and the fact that the components of $G^\subl$ are balanced, $\ket{0}_\subl$ is the unique ground state of $\Hham\pap{G^\subl}$   with $S=0$ and we can define the energy gap $\Delta^\subl:=\lambda^\subl_1-\lambda^\subl_0 >0$. 
Similarly, $\ket{k,\beta}_\supl$ are the eigenstates with energy $\lambda^\supl_k(\beta)$ of $\Hham\pap{G^\supl_\beta}$ with $k\in\pas{0,2^{|V|^\supl}-1}\subset \mathbb N$. 
The unbalanced component of $G^\supl_\beta$ implies the ground states of the Hamiltonian $\Hham\pap{G^\supl_\beta}$ have an $S$ given by Theorem 1. 
Therefore, the ground states have energy $\lambda_0^\supl(\beta)=\lambda_m^\supl(\beta)\;\forall m\in\pas{0,2S}\subset \mathbb N$ and we can define the gap $\Delta^\supl(\beta):=\lambda_{2S+1}^\supl(\beta)-\lambda_0^\supl(\beta)$.

Now, we assume that $\ket{0}_\subl$ is superstable with respect to all $\mathcal{X}_{i,q}$. It implies $\ket{0,k,\beta}$ is superstable to the interactions in Eq. \eqref{Eq_GEN_SSProblem_NOSUBG} $\forall k, \beta$. 
Consider the subset of superstable states given by $\ket{0,m,\beta}$ with $m$ in the abovementioned range. We want to show that there exists a range of positive $\beta$ such that the $\ket{0,m,\beta}$ are the only ground states for Eq. \eqref{Eq_GEN_SSProblem_NOSUBG} and that these are gapped from the rest of the spectrum. These states have a total spin $S$ as defined in Theorem 1, respecting the trivial degeneracy. 

 The states $\ket{0,m,\beta}$ have eigenvalues $E^\sub_0(\beta):=\lambda^\subl_0+\lambda_0^\supl(\beta)$, and 
we require $E^\sub_0(\beta)<\bra{\psi}H^\sub(\beta)\ket{\psi}$ for all other eigenstates of $H^\sub(\beta)$.
We start by comparing the other superstable states $\ket{0,k>2S,\beta}$. Their energy is bounded from below as 
 \begin{equation*}
\bra{0,k>2S,\beta}H^\sub(\beta)\ket{0,k>2S,\beta}\geq E^\sub_0(\beta)+ \Delta^\supl(\beta).
 \end{equation*}
This lower bound is always larger than $E^\sub_0(\beta)$. Hence, none of the other superstable states can feature equal or lower eigenenergies than $E^\sub_0(\beta)$.
All the other eigenstates orthogonal to $\ket{0,k,\beta}\forall k$ may be expanded as
\begin{equation*}
\ket{\psi}=\sum_{j=1}^{2^{|V|^\subl}-1}\sum_{k=0}^{2^{|V|^\supl}-1}C_{j,k}\ket{j,k,\beta}.
\end{equation*}
It is easy to show that any state of this form provides a lower bound for the energy as
\begin{equation*}
    \bra{\psi}H^\sub(\beta)\ket{\psi}\geq E^\sub_0(\beta) +\Delta^\subl -\sum_{i=1}^{(n+1)/2}\beta^i \overline{Q} \overline{\mathcal{X}}  \overline{\mathcal{Y}},
\end{equation*}
where $\overline{Q}:=\max_i(Q_i)$, $\overline{\mathcal{X}}:=\max_{i,q}(||\mathcal{X}_{i,q}||)$, and $\overline{\mathcal{Y}}:=\max_{i,q}(||\mathcal{Y}_{i,q}||)$; with $||O||$ denoting the operator norm of the operator $O$.  Consequently,
\begin{equation}
    \beta \frac{1-\beta^{(n+1)/2}}{1-\beta} \overline{Q} \overline{\mathcal{X}}  \overline{\mathcal{Y}}< \Delta^\subl
\end{equation}
provides a sufficient condition so that $\ket{0,m,\beta}$ are the gapped ground states for finite $\beta$. In other words, this inequality determines the regime of sufficiently small $\beta$ where the ground states are only the superstable states $\ket{0,m,\beta}$. 

Therefore, for sufficiently small $\beta$, the ground states of a Hamiltonian proportional to Eq. \eqref{Eq_GEN_SSProblem_NOSUBG} are exactly $(2S+1)$ degenerate, each being the unique ground state with a different quantum number for the spin projection along the $z$-axis. Any ground state that can be adiabatically continued to one of these while preserving the {\rm SU}(2) symmetry must have the same total spin quantum number $S$. This establishes that the ground state of the Hubbard model on a superstable graph has total spin given by $S$ in the strongly correlated regime for sufficiently small $\beta$. $\blacksquare$ 

\subsection{Proof of Corollary 1}
\label{Sec-ProofCor1}
For sufficiently large $U$ and sufficiently small $\beta$, the ground states of the Hamiltonian obtained by the order $n$ expansion are given by 
\begin{equation}
\label{Eq_GEN_SSGS}
    \ket{0,m,\beta}=\bigotimes_{u\in \mathcal{I}}\ket{0}_{\subl_u}\bigotimes_{v\in \mathcal{I}|v\neq 0}\ket{0,\beta}_{\supl_v}\otimes \ket{m,\beta}_{\supl_0},
\end{equation}
with $\ket{0}_{\subl_u}$ the ground state of $\Hham(G^\subl\pas{\subl_u})$. Here, $\ket{k,\beta}_{\supl_v}$ is the $k$-th eigenvector ordered ascendingly in the eigenvalues of  $\Hham(G^\supl_\beta\pas{\supl_v})$ and $\supl_0$ is the set of vertices corresponding to the only unbalanced component of $G^\supl_\beta$.
This product form of the wavefunction ensures the correlations between the spaces spanned by the different components vanish. Within each component of $G^\subl\oplus G^\supl_\beta$, the Shen-Qiu-Tian theorem gives the sign of the ground state correlations of the effective model. As shown in \cite{supplementary_material}, the $\expv{S^+_iS^-_j}$ correlations of the effective model obtained through the SW transformation represent the correlations of the original system up to $O(U^{-2})$ corrections. $\blacksquare$
\newpage 
\appendix
\section*{Supplementary Material For: Magnetization of electronic ground states in frustrated superstable graphs} 
\section{Schrieffer-Wolff transformation and low energy physics}
\label{APP:SWT}
The Schrieffer-Wolff (SW) transformation is a unitary transformation, $H'(G):=e^WH(G)e^{-W}$, yielding a block-diagonal Hamiltonian.
In the case of the Hubbard model, the  SW transformation is conventionally used to derive
effective low-energy Hamiltonians where the doublon number - i.e., number of doubly occupied sites - is a good quantum number \cite{t-U_Expansion}. To do this, we rewrite the Hubbard Hamiltonian as $H(G)=V+T$, with $V:=U\sum_{i\in V}n_{i,\uparrow}n_{i,\downarrow}$, and $T:=-\sum_{\pap{i,j} \in E,\sigma}t_{i,j}c_{i,\sigma}^\dagger c_{j,\sigma}$, where we use $t_{i,j}:=t(e)$ for $e=(i,j)$ to denote the weights. Using the  Baker–Campbell–Hausdorff formula, we obtain
\begin{equation}
\label{eq:UnitaryTransformationOfH}
    H'(G)=H(G)+[W,H(G)]+\frac{1}{2}[W,[W,H(G)]]+...,
\end{equation}
where the operator $W$ is chosen
such that $[V,H'(G)]=0$. The expansion series in Eq. \eqref{eq:UnitaryTransformationOfH} is absolute convergent when $U>16||T||$ with $||T||$ denoting the operator norm of $T$ \cite{BRAVYI_SWT}.

Next, we define the hole number operator of spin $\sigma$ $h_{i,\sigma}:=1-n_{i,\sigma}$; and the $\sigma$-dependent hopping from $i$ to $j$, $T_{i,j,\sigma}:=-t_{i,j}c_{i,\sigma}^\dagger c_{j,\sigma}$. Clearly, $1=h_{i,\sigma}+n_{i,\sigma}$. With this, we replace $1=h_{i,\Bar{\sigma}}+n_{i,\Bar{\sigma}}$ at the left and $1=h_{j,\Bar{\sigma}}+n_{j,\Bar{\sigma}}$ at the right of $T$, where $\Bar{\sigma}$ is the opposite of the spin direction $\sigma$. This splits the hopping into three terms $T=T_0+T_1+T_{-1}$, with
\begin{equation}
T_0:=\sum_{i,j,\sigma}\pap{n_{i,\Bar{\sigma}}T_{i,j,\sigma}n_{j,\Bar{\sigma}}+h_{i,\Bar{\sigma}}T_{i,j,\sigma}h_{j,\Bar{\sigma}}};
\end{equation}
\begin{equation}
T_{1}:=\sum_{i,j,\sigma}n_{i,\Bar{\sigma}}T_{i,j,\sigma}h_{j,\Bar{\sigma}};\quad
        T_{-1}:=\sum_{i,j,\sigma}h_{i,\Bar{\sigma}}T_{i,j,\sigma}n_{j,\Bar{\sigma}}.
\end{equation}
$T_0$ does not change the doublon number of a state. Instead, $T_1$ and $T_{-1}$ increase or decrease the doublon number by $1$, respectively. It is easy to show that $T_m^\dagger=T_{-m}$ and
\begin{equation}
\label{eq:Tm_V_commutator}
    [V,T_m]=mUT_m,
\end{equation}
with $m\in\{-1,0,1\}$. Let us also define the products $T(m_1,m_2,...,m_n):=T_{m_1}T_{m_2}...T_{m_n}$, $m_i \in \{-1,0,1\}$. It can be shown by induction that
\begin{equation}
\label{eq:Commu_UT_Many}    [V,T(m_1,m_2,...,m_n)]=\pap{\sum_{i=1}^n m_i}UT(m_1,m_2,...,m_n).
\end{equation}
From Eq. \eqref{eq:UnitaryTransformationOfH}, we see that, at the first order, $W$ has to cancel the $T_m$'s with $|m|=1$ in the Hamiltonian~$H(G)$ to obtain a well defined doublon number. From Eq. \eqref{eq:Tm_V_commutator}, it is clear that $W$ is a linear combination of  $T_m$'s with $|m|=1$. For higher orders, the terms to cancel are linear combinations of different $T(m_1,m_2,...,m_n)$ with non-zero $\sum_{i=1}^n m_i$. Similarly, Eq. \eqref{eq:Commu_UT_Many} shows that higher orders of $W$ are also linear combinations of $T(m_1,m_2,...,m_n)$. Note that $[T_m,S^\alpha]=0\forall m$ implies $[T(m_1,m_2,...,m_n),S^\alpha]=0$. Hence, the  SW transformation preserves the total spin and the global spin rotation invariance.  Furthermore, the $T_m$'s are time reversal invariant; consequently, the transformation also preserves this symmetry.

Up to the third order, the Hamiltonian resulting from the SW transformation reads 
\begin{widetext}
 \begin{equation}
\begin{split}
 H'(G)=&V+T_0+\frac{1}{U}\pac{T(1,-1)-T(-1,1)}\\      
 &+\frac{1}{U^2}\{T(-1,0,1)+T(1,0,-1)\\ &-\frac{1}{2}\pas{T(0,-1,1)+T(0,1,-1)+T(-1,1,0)+T(1,-1,0)}\}\\
 &+\frac{1}{U^3}\{T(0,-1,0,1)+T(-1,0,1,0)+T(-1,1,-1,1)+T(1,0,0,-1)\\
 &-\pas{T(0,1,0,-1)+T(-1,0,0,1)+T(1,-1,1,-1)+T(1,0,-1,0)}\\
 &+\frac{1}{2}[T(0,0,1,-1)+T(1,-1,0,0)+T(1,1,-1,-1)\\ &-\pap{T(0,0,-1,1)+T(-1,-1,1,1)+T(-1,1,0,0)} ] \}+O(U^{-4}).
\end{split}
\end{equation}
\end{widetext}
With large $U$ and half filling, the ground state of Eq. \eqref{eq:UnitaryTransformationOfH} belongs to the $0$-doublon sector. Thus, only terms with a $T_1$ operator on the right contribute to the low-energy physics. In the $0$-doublon sector, the Hamiltonian can be written in terms of spin $1/2$ operators. Due to spin rotation and time reversal invariances, the terms with even order in $1/U$ vanish (see below), leading to
\begin{equation}
\label{Eq:LargeUThirdOrder}
\begin{split}
       H'(G)=&\Hham\pap{G_{\rm H}}+\frac{1}{U}\Hham\pap{G_{\rm H}}^2\\
       &-\frac{1}{U^3}\pas{T(-1,0,0,1)
+\frac{1}{2}T(-1,-1,1,1)}\\
  &+O(U^{-5}),  
\end{split}
 \end{equation}
 where $\Hham\pap{G}$ and $G_{\rm H}$ are the Heisenberg Hamiltonian and the graph $G$ with renormalized weights as defined in the Main Text (MT). 
\subsection*{Transformed magnetic correlations}
\label{APP_SWTMagneticCorrs}
The ground state magnetic correlations of the Hubbard model, $\expv{S^+_iS^-_j}$ are expressed in terms of the ground state of the transformed system as
\begin{equation}
    \expv{S^+_iS^-_j}=\expv{e^WS^+_iS^-_je^{-W}}_{\rm T},
\end{equation}
where $\expv{\cdot}_{\rm T}$ takes the expected value with respect to the ground state of $H'(G)$. Using $\pas{V,S^+_iS^-_j}=0$; the Baker-Campbell-Hausdorff formula; and the fact that $W$ is strictly an off-diagonal operator connecting sectors with different doublon number \cite{BRAVYI_SWT}, one obtains
\begin{equation}
    \expv{S^+_iS^-_j}=\expv{S^+_iS^-_j}_{\rm T}+O(U^{-2}).
\end{equation}
\section{General spin rotation and time reversal invariant Hamiltonian}
\label{App:TRSR_Symmetries}
In this section, we derive conditions for the most general spin-$1/2$ Hamiltonian composed of $|V|$ sites, labeled from 1 to $|V|$, that is invariant under global spin rotations. We use the form of this Hamiltonian to express the effective  SW transformation Hamiltonian in Eq. \eqref{Eq:LargeUThirdOrder} in terms of spin operators, where factorization into the form of a $H^\sub(\beta)$ Hamiltonian is more straightforward. Consider a Hamiltonian $H_N$ with $N\leq |V|$ that consists  of interaction terms involving products of $N$ non-identity operators acting on $N$ different sites. Its most general form is 
\begin{equation}
H_N=\sum_{J\in\pac{P(I)}_N}a^{J_1,J_2,...,J_N}_{\alpha_1,\alpha_2,...,\alpha_N}S^{\alpha_1}_{J_1}S^{\alpha_2}_{J_2}...S^{\alpha_N}_{J_N},
\end{equation}
where $S^\alpha$ with $\alpha\in\pac{1,2,3}$ are the spin operators in the $x$, $y$, and $z$ directions, respectively; $I:=[1,|V|]\subset \mathbb{N}$ is the set with all the site labels; $P(I)$ is the power set of $I$; $\pac{A}_N$ selects the subsets with cardinality $N$ in the set of sets $A$; and $J_i$ are the components of the $N-$tuple of site indices $J$, which specify the sites interacting for the given term. We use the sum convention in the $\alpha_i \in{\pac{1,2,3}}$ indices and simplify the notation by defining the $N-$tuple $\bm{\alpha}:=\pac{\alpha_1,\alpha_2,...,\alpha_N}$. Thus,
\begin{equation}
\label{Eq:GeneralNHam}
   H_N= \sum_{J\in\pac{P(I)}_N}a^{J}_{\bm{\alpha}}\prod_{i=1}^N S^{\alpha_i}_{J_i},
\end{equation}
where $a^{J}_{\bm{\alpha}}\in \mathbb{R}$ guarantees that the Hamiltonian is Hermitian. We recall that a global spin rotation invariant Hamiltonian must commute with the generators of the symmetry group, i.e., $S^\beta\forall\beta \in \pac{1,2,3}$. Hence, we impose rotational invariance by requiring that $a^{J}_{\bm{\alpha}}$ yields to $\pas{H_N,S^\beta}=0 \forall \beta$. It follows that 
\begin{equation}
i\sum_{J\in\pac{P(I)}_N}\sum_{i=1}^Na^{J}_{\bm{\alpha}}\epsilon_{\alpha_i,\beta,\gamma}S^{\gamma}_{J_i}\prod_{j=1,j\neq i}^N S^{\alpha_j}_{J_j}=0,
\end{equation}
where we have used the commutator $\pas{S^{\alpha}_{i},S^{\beta}_{j}}=i\epsilon_{\alpha,\beta,\gamma}\delta^{i,j}S^{\gamma}_{i}$, with $\epsilon_{\alpha,\beta,\gamma}$ the Levi-Civita symbol. Due to the implicit summation in the direction of the spin operators, we can interchange the labels $\gamma\leftrightarrow \alpha_i$. Furthermore, let us define 

\[
\bm{\alpha}\pap{i,j} :=
\begin{cases}
  \{\alpha_i, \alpha_{i+1}, \ldots, \alpha_{j-1}, \alpha_j\} & \text{if } i \leq j, \\
  \emptyset & \text{otherwise}.
\end{cases}
\]
Then the equation reads
\begin{equation}
        i\sum_{J\in\pac{P(I)}_N}\sum_{i=1}^Na^{J}_{\bm{\alpha}\pap{1,i-1}\cup\pac{\gamma}\cup\bm{\alpha}\pap{i+1,N}}\epsilon_{\gamma,\beta,\alpha_i}\prod_{j=1}^N S^{\alpha_j}_{J_j}=0.
\end{equation}
The spin operators are linearly independent. Therefore, for a given $J\in\pac{P(I)}_N$, $a^{J}_{\bm{\alpha}}$ must follow
\begin{equation}
  \sum_{i=1}^N  a^{J}_{\bm{\alpha}\pap{1,i-1}\cup\pac{\gamma}\cup\bm{\alpha}\pap{i+1,N}}\epsilon_{\alpha_i,\beta,\gamma}=0.
\end{equation}
For $N=1$, the equation has only the trivial solution. For $N=2$ the solution is $a^{J_1,J_2}_{\alpha_1,\alpha_2}=a^{J_1,J_2}\delta_{\alpha_1,\alpha_2}$ results in a Hamiltonian that is formed by the dot product between the spin operators of the sites $J_1$ and $J_2$ times a constant. Therefore, the Heisenberg Hamiltonian is the only possible spin rotation invariant Hamiltonian with $N=2$. For $N=3$, the result is a Hamiltonian of the form $a^{J_1,J_2,J_3}S_{J_1}\cdot\pap{S_{J_2}\times S_{J_3}}$. For $N=4$, one obtains three solutions that correspond to the three permutations that produce a product of two dot products with four different sites. 

We further restrict the $a^{J}_{\bm{\alpha}}$ by introducing time reversal symmetry. For the Hamiltonian in Eq. \eqref{Eq:GeneralNHam}, time reversal amounts to the transformation $S^\alpha_i\rightarrow -S^\alpha_i$. Hence, the $H_N$'s with odd $N$ are forbidden. From these results, we conclude that the most general spin Hamiltonian of $|V|$ spins is
\begin{equation}
\label{Eq_GenHamWithV}
H=\sum_{i=0}^{\lfloor |V|/2 \rfloor} H_{2i},    
\end{equation}
where $H_0=a$ with $a$ a constant.
\subsection*{Third-order spin Hamiltonian}
\label{APP_FactorizationForClusters}
Hence, the $n=3$ expansion given in Eq. \eqref{Eq:LargeUThirdOrder} has the form of Eq. \eqref{Eq_GenHamWithV} with $H_{2i}=0$ for $i>2$. This leads to the spin model 
\begin{equation}
\label{Eq-ThirdOSpin}
  H_3(G):=\Hham\pap{G_{\rm H}}+H_{\rm R}(G)+H_{\rm TP}\pap{G}+H_{\rm FC}(G).  
\end{equation}
Here, the second term is a renormalization of the Heisenberg terms
\begin{equation}
    H_{\rm R}(G):=\frac{16}{U^3}\sum_{e=\pap{i,j} \in E }t\pap{e}^4 \pap{\frac{1}{4}-\Vec{S}_i \cdot \Vec{S}_j}.
\end{equation}
The third term is given by
\begin{equation}
H_{\rm TP}\pap{G}:=\frac{4}{U^3}\sum_{p \in {\rm TP}(G)} \pas{t_{p,\pap{0,1}}t_{p,\pap{1,2}}}^2\pas{\Vec{S}_{p,0}\cdot \Vec{S}_{p,2}-\frac{1}{4}}.
\end{equation}
The set ${\rm TP}(G)$ comprises the indices $p$ that label the simple two paths of $G$. Any vertex $i$ part of a two path can be identified by the tuple $(p,\so)$, where $\so\in\pac{0,1,2}$ labels the vertex in the path. $1$ is the label for the site connecting the two ends $0,2$. Thus, the edges in the simple two path  are $\pap{(p,0),(p,1)}$ and $\pap{(p,1),(p,2)}$. In this notation, the path index and a pair of sites within the path determine the weights $t_{p, (i,j)}$. 

The fourth term contains the four cycles, 
\begin{widetext}
\begin{equation}
\begin{split}
    H_{\rm FC}(G)&:=\frac{4}{U^3}\sum_{l\in {\rm FC}(G) } \Bigg\{ t_{l,\pap{0,1}}t_{l,\pap{0,2}}t_{l,\pap{1,3}}t_{l,\pap{2,3}}\bigg\{   \frac{1}{4}\\ 
    &-\pas{\pap{\Vec{S}_{l,0}\cdot \Vec{S}_{l,1}}+\pap{\Vec{S}_{l,0}\cdot \Vec{S}_{l,2}}+\pap{\Vec{S}_{l,0}\cdot \Vec{S}_{l,3}}+\pap{\Vec{S}_{l,1}\cdot \Vec{S}_{l,2}}+\pap{\Vec{S}_{l,1}\cdot \Vec{S}_{l,3}}+\pap{\Vec{S}_{l,2}\cdot \Vec{S}_{l,3}}}\\
    &+20\pas{\pap{\Vec{S}_{l,0}\cdot \Vec{S}_{l,1}}\pap{\Vec{S}_{l,2}\cdot \Vec{S}_{l,3}}+ \pap{\Vec{S}_{l,0}\cdot \Vec{S}_{l,2}}\pap{\Vec{S}_{l,1}\cdot \Vec{S}_{l,3}}-\pap{\Vec{S}_{l,0}\cdot \Vec{S}_{l,3}}\pap{\Vec{S}_{l,1}\cdot \Vec{S}_{l,2}}}\bigg\}  \Bigg\}.
\end{split}
\end{equation}
\end{widetext}
Similar to ${\rm TP}(G)$, the set ${\rm FC}(G)$ is the set of the indices labeling each simple four cycle of $G$. Any vertex $i$ can be identified with the cycle index $l\in {\rm FC}(G)$ and the vertex index within the cycle $\sil \in\pac{0,1,2,3}$ through the $(l,\sil)$ tuple. The weights, $t_{l, (i,j)}$, are defined analogously to the simple two path case.

\section{Superstability up to the third-order expansion}
\label{APP_FactorizationForClusters}
Here, we demonstrate that superstable graphs detected by our algorithm (see EM) have an associated spin Hamiltonian, up to third order [cf. Eq. \eqref{Eq-ThirdOSpin}], with the form of a superstable spin Hamiltonian $H^\sub(\beta)$. We start from the graph partition in Eq. (C1) of the EM. 
To simplify the notation in what follows, we define $V^\subl:=\sqcup_{u\in \mathcal{I_\subl}}\subl_u$, $E^\subl:=\sqcup_{u\in \mathcal{I_\subl}}E^{\subl_u}$, $t_\subl:=\min_{e\in E^\subl} [|t(e)|]$, $t_\supl:=\max_{e\in E \setminus E^\subl} [|t(e)|]$, $t^\subl (e):= t(e)/t_\subl \forall e \in E^\subl$, and $t^\supl (e):= t(e)/t_\supl \forall e \in E^\supl \sqcup E^\Li$.
Hence we have, $\beta=(t_\supl/t_\subl)^2$, $|t^\subl (e)|\geq 1$ and $|t^\supl (e)|\leq 1$. 
Additionally, we define the parameter $\alpha:=t_\subl/U$ to keep track of the order of the terms. 

\subsection{Superstable graphs identified by the first expansion order}
\label{Sec_Case1}
Consider the case of superstable graphs for which $G\pas{V^\supl}$ is bipartite with at most one unbalanced connected component. Their $n=1$  SW transformation expansion is given by $\Hham\pap{G_{\rm H}}$. The uniformity of the connection between cores and residues, and the algebraic relation describing the weights connecting cores (see EM) allows us to factorize the Hamiltonian as
\begin{widetext}
\begin{equation}
\label{Eq_Case1SS}
\begin{split}
    \Hham\pap{G_{\rm H}}&=4\alpha t_\subl \Bigg \{\Hham\pap{G^\subl}+ \Hham\pap{G^\supl_\beta}+\beta \sum_{y\in V^\supl,u\in \mathcal{I}_\subl} t^\prime_u(y)^2  \pap{\Vec{S}_{u} \cdot \Vec{S}_y-\frac{|\subl_u|}{4}}\\
    &+\beta \sum_{u,v>u|\pac{u,v}\in \mathcal{I}_\subl}\pas{ \sum_{x\in \subl_u} t^\prime_{\subl_u-\subl_v}(x)^2  \pap{\Vec{S}_{v} \cdot \Vec{S}_x-\frac{|\subl_v|}{4}}+\sum_{x^\prime\in \subl_v} t^\prime_{\subl_v-\subl_u}(x)^2\pap{  \Vec{S}_{u} \cdot \Vec{S}_{x^\prime}-\frac{|\subl_u|}{4}}} \Bigg \}.
\end{split}
\end{equation}
\end{widetext}
In this way, we have factorized constants to tune the Hamiltonian with dimensionless parameters $\alpha$ and $\beta$. The terms in the curly brackets have the form of a $H^\sub(\beta)$ Hamiltonian. The first two terms are Heisenberg models on the $G^\subl:=(V^\subl,E^\subl,t^\subl(E^\subl)^2)$ and $G^\supl_\beta:=(V^\supl,E^\supl,\beta t^\supl(E^\supl)^2)$ graphs, which, respectively, have the same edges and vertices of $\bigoplus_{u\in \mathcal{I_\subl}} G\pas{\subl_u}$ and $G_{\rm }\pas{V^\supl}$ but their weights are transformed. 
The remaining terms come from the frustration in the system. The third term in the first line comes from the interactions between cores and residue, while the last line describes interactions between cores. Here, the primed weights denote that they have been divided by $t_\supl$. All the terms coming from frustration contain the total spin operator of a core labeled with index $u$, $\Vec{S}_{u}:=\sum_{x\in \subl_u}\Vec{S}_{x}$. Since $G\pas{\subl_u}$ is balanced $\forall u$, the ground state of $\Hham\pap{G^\subl}$, $\bigotimes_{u\in \mathcal{I}_\subl}\ket{0}_{\subl_u}$, leads to $S_{u}^\alpha \bigotimes_{v\in \mathcal{I}}\ket{0}_{\subl_v}=0$. Thus, the form of the frustrating terms in Eq. \eqref{Eq_Case1SS} guarantees that $\ket{0}_\subl$ is superstable with respect to all $\mathcal{X}_{i,q}$, which is required in the proof of our theorem to ensure that the ground state magnetization is unique. By the assumption on the connected components of $G\pas{V^\supl}$, the graph $G^\supl_\beta$ is bipartite with at most one unbalanced connected component. Hence, a graph with the properties identified by our algorithm yield a $H^\sub(\beta)$ Hamiltonian in the $n=1$ order. 

\subsection{Superstable graphs identified by the third expansion order}
\label{Appendix_Case2}
When the induced graph $G_{\rm }\pas{V^\supl}$ has more than one unbalanced component, we obtain multiple degenerate superstable ground states at $n=1$. This degeneracy breaks in the next order for the superstable graphs characterized by the $n=3$  SW transformation expansion order, given by the additional restrictions in the EM. Here, we show how the additional terms emerging from this expansion order generate the form of $H^\sub(\beta)$ and can add new edges between the components.

The Eq. \eqref{Eq-ThirdOSpin} contains a Heisenberg term coming from the first order of the expansion. This term is factorized as in Eq. \eqref{Eq_Case1SS}. However, the graph coming from $G_{\rm }\pas{V^\supl}$ is not $G^\supl_\beta$ for this third-order expansion, as the latter graph has, by definition, at most one unbalanced component. Thus, to obtain the correct contribution of the third-order expansion, $G^\supl_\beta$ must be replaced by $G^{\supl \prime}_\beta:=(V^\supl,E^\supl,\beta t^\supl(E^\supl)^2)$ in Eq. \eqref{Eq_Case1SS}, which has more than one unbalanced component. The graph $G^{\supl \prime}_\beta$ serves as a structure accommodating multiple unbalanced components from which the graph $G^\supl_\beta$ is obtained by adding edges coming from the higher order terms. Analogous to the definition of the $\supl_v$ sets in the EM, we define $G^{\supl \prime}_\beta=\bigoplus_{v\in \mathcal{I}}  G^{\supl \prime}_\beta\pas{\supl^\prime_v}$. Here, the set of vertices
$\supl^\prime_v\subseteq V^\supl$ also collectively form the residue vertex set $V^\supl=\bigsqcup_{v\in \mathcal{I}}\supl^\prime_v$. 

In addition to the $n=1$ terms, the $n=3$ expansion contains a renormalization term, two path and four cycle terms. We absorb the renormalization term, $H_{\rm R}(G)$, in $\Hham\pap{G_{\rm H}}$, which corresponds to subtracting $16t(e)^4/U^3$ to each weight of $G_{\rm H}$. While the term changes the weights of the graph, it does not add new edges or beyond-pairwise interactions. In contrast, the two path and four cycle terms can induce new couplings between components. Our goal is to show that the resulting terms either generate interaction terms that include an $S_{u}^\alpha$ operator or that they connect vertices of different $\supl^\prime_v$ in a way that allows us to define a graph $G^\supl_\beta$ with a unique unbalanced component. 

First, we focus on the terms emerging from the two paths. We categorize the interacting terms based on the components involved in the interaction. We use the notation $A-B-C$ to denote the two paths that start at the component $A$, then move to a vertex in the component $B$, and finally end in a vertex of the component $C$. The possible terms emerging from the two paths are:


\begin{widetext}
\begin{itemize}
    \item {\bf $\subl_u-\supl^\prime_v-\subl_w$. } 
        \begin{equation}
        \label{Eq_SS_TP1}
        \frac{4}{U^3}\sum_{u,v,w>u,x\in \subl_u, y\in \supl^\prime_v,x^\prime \in \subl_w } \pas{t_{x,y}t_{y,x^\prime}}^2\pas{\Vec{S}_{x}\cdot \Vec{S}_{x^\prime}-\frac{1}{4}}=\frac{4}{U^3}\sum_{u,v,w>u, y\in \supl^\prime_v } \pas{t_{\subl_u}(y)t_{\subl_w}(y)}^2\pas{\Vec{S}_{u}\cdot \Vec{S}_{w}-\frac{|\subl_u||\subl_w|}{4}}.
    \end{equation}
    \item \bf $\subl_u-\supl^\prime_v-\subl_u$ .
    \begin{equation}
    \label{Eq_SS_TP1-2}
                \frac{4}{U^3}\sum_{u,v,x,x^\prime>x | \pac{x,x^\prime}\in \subl_u, y\in \supl^\prime_v } \pas{t_{x,y}t_{y,x^\prime}}^2\pas{\Vec{S}_{x}\cdot \Vec{S}_{x^\prime}-\frac{1}{4}}=\frac{4}{U^3}\sum_{u,v, y\in \supl^\prime_v} \frac{t_{\subl_u}(y)^4}{2}\pas{\Vec{S}_{u}\cdot \Vec{S}_{u}-\frac{|\subl_u|}{4}\pap{|\subl_u|+2}}.
    \end{equation}
    \item  {\bf $\subl_u-\subl_u-\supl^\prime_v$. }
        \begin{equation}
        \label{Eq_SS_TP2}
        \frac{4}{U^3}\sum_{u,v,x,x^\prime\neq x|\pac{x,x^\prime}\in \subl_u, y\in  \supl^\prime_v } \pas{t_{x,x^\prime}t_{x^\prime,y}}^2\pas{\Vec{S}_{x}\cdot \Vec{S}_{y}-\frac{1}{4}}=\frac{4}{U^3}\sum_{u,v, y\in \supl^\prime_v } \pas{t_{x,x^\prime}t_{\subl_u}(y)}^2\pas{\pap{|\subl_u|-1}\Vec{S}_{u}\cdot \Vec{S}_{y}-\frac{|\subl_u|}{4}\pap{|\subl_u|-1}}.
    \end{equation}
        \item  {\bf $\subl_u-\supl^\prime_v-\supl^\prime_v$. }
        \begin{equation}
        \label{Eq_SS_TP3}
        \frac{4}{U^3}\sum_{u,v,x\in\subl_u, y,y^\prime|\pac{y,y^\prime}\in\supl^\prime_v}  \pas{t_{x,y}t_{y,y^\prime}}^2\pas{\Vec{S}_{x}\cdot \Vec{S}_{y^\prime}-\frac{1}{4}}=\frac{4}{U^3}\sum_{ u,v,y,y^\prime|\pac{y,y^\prime}\in\supl^\prime_v } \pas{t_{\subl_u}(y)t_{y,y^\prime}}^2\pas{\Vec{S}_{u}\cdot \Vec{S}_{y^\prime}-\frac{|\subl_u|}{4}}.
    \end{equation}

\item  {\bf $\subl_u-\subl_u-\subl_v$.}
        \begin{equation}
        \label{Eq_SS_TPC1}
        \frac{4}{U^3}\sum_{u,v\neq u, y,y^\prime|\pac{y,y^\prime}\in\subl_u,x\in\subl_v}  \pas{t_{y,y^\prime}t_{y^\prime,x}}^2\pas{\Vec{S}_{y}\cdot \Vec{S}_{x}-\frac{1}{4}}=\frac{4}{U^3}\sum_{ u,v\neq u, y,y^\prime|\pac{y,y^\prime}\in\subl_u} \pas{t_{\subl_u-\subl_v}t_{y,y^\prime}}^2\pas{\Vec{S}_{v}\cdot \Vec{S}_{y}-\frac{|\subl_v|}{4}}.
    \end{equation}

\item  {\bf $\subl_u-\subl_v-\subl_u$.  }
        \begin{equation}
        \label{Eq_SS_TPC3}
    \frac{4}{U^3}\sum_{u,v\neq u,x,x^\prime>x | \pac{x,x^\prime}\in \subl_u, y\in \subl_v } \pas{t_{x,y}t_{y,x^\prime}}^2\pas{\Vec{S}_{x}\cdot \Vec{S}_{x^\prime}-\frac{1}{4}}=\frac{4}{U^3}\sum_{u,v\neq u, y\in \subl_v} \frac{t_{\subl_u-\subl_v}^4}{2}\pas{\Vec{S}_{u}\cdot \Vec{S}_{u}-\frac{|\subl_u|}{4}\pap{|\subl_u|+2}}.
    \end{equation}

\item  {\bf $\subl_u-\subl_v-\subl_w$. }
        \begin{equation}
        \label{Eq_SS_TPC4}
\frac{4}{U^3}\sum_{u,v\neq u,w>u|w\neq v,x\in \subl_u, y\in \subl_v,x^\prime \in \subl_w } \pas{t_{x,y}t_{y,x^\prime}}^2\pas{\Vec{S}_{x}\cdot \Vec{S}_{x^\prime}-\frac{1}{4}}=\frac{4}{U^3}\sum_{u,v\neq u,w>u|w\neq v, y\in \subl_v } \pas{t_{\subl_u-\subl_v}t_{\subl_v-\subl_w}}^2\pas{\Vec{S}_{u}\cdot \Vec{S}_{w}-\frac{|\subl_u||\subl_w|}{4}}.
    \end{equation}

    \item  {\bf $\subl_u-\subl_v-\supl^\prime_w$. }
        \begin{equation}
        \label{Eq_SS_TPC5}
 \frac{4}{U^3}\sum_{u ,v\neq u, w,x\in \subl_u, x^\prime\in \subl_v, y\in \supl_w^\prime } \pas{t_{x,x^\prime}t_{x^\prime,y}}^2\pas{\Vec{S}_{x}\cdot \Vec{S}_{y}-\frac{1}{4}}=\frac{4}{U^3}\sum_{u,v\neq u,w, x^\prime\in \subl_v, y\in \supl_w^\prime } \pas{t_{\subl_u-\subl_v}t_{\subl_v}(y)}^2\pas{\Vec{S}_{u}\cdot \Vec{S}_{y}-\frac{|\subl_u|}{4}}.
    \end{equation}

                \item  {\bf $\supl^\prime_u-\subl_v-\supl^\prime_u$. }
        \begin{equation}
        \label{Eq_SS_TPC9}
\frac{4}{U^3}\sum_{u,v,y,y^\prime>y|\pac{y,y^\prime}\in\supl^\prime_u,x\in\subl_v } \pas{t_{y,x}t_{x,y^\prime}}^2\pas{\Vec{S}_{y}\cdot \Vec{S}_{y^\prime}-\frac{1}{4}}=\frac{4}{U^3}\sum_{u,v,y,y^\prime>y|\pac{y,y^\prime}\in\supl^\prime_u } \pas{t_{\subl_v}(y)t_{\subl_v}(y^\prime)}^2|\subl_v|\pas{\Vec{S}_{y}\cdot \Vec{S}_{y^\prime}-\frac{1}{4}}.
    \end{equation}

        \item  {\bf $\supl^\prime_u-\subl_v-\supl^\prime_w$.}
        \begin{equation}
        \label{Eq_Heisenberg_TP}
        \frac{4}{U^3}\sum_{u,v,w>u,y\in\supl^\prime_u,x\in\subl_v, y^\prime\in\supl^\prime_w  } \pas{t_{y,x}t_{x,y^\prime}}^2\pas{\Vec{S}_{y}\cdot \Vec{S}_{y^\prime}-\frac{1}{4}}=\frac{4}{U^3}\sum_{u,v,w>u,y\in\supl^\prime_u, y^\prime\in\supl^\prime_w } \pas{t_{\subl_v}(y)t_{\subl_v}(y^\prime)}^2|\subl_v|\pas{\Vec{S}_{y}\cdot \Vec{S}_{y^\prime}-\frac{1}{4}}.
    \end{equation}
\end{itemize}
\end{widetext}
On the right-hand side of the equations, we have used the weight properties of the connections with the cores of the superstable graphs, as described in the EM. To express these sums, we adopt an indexing convention  where core components are assigned smaller integers than the residue components. Additionally, vertex labels inherit the ordering from the labeling of the components. To formalize this, we define $G_{\rm S}:=G^\subl\oplus G^{\supl \prime}_\beta$, representing the effective spin model graph at first order. The connected  components of $G_{\rm S}$ are indexed as $G_{\rm S}=\bigoplus_{u=1}^{n_{\subl}}  G^\subl\pas{\subl_u}\bigoplus_{v=1+n_{\subl}}^{n_{\subl}+n_{\supl}}  G^{\supl \prime}_\beta\pas{\supl_v}$, where $n_{\subl}$ and $n_{\supl}$ denote the number of components of $G^\subl$ and $G^{\supl \prime}_\beta$, respectively. The vertex indexing for the cores is given by: $x\in \subl_1 \Rightarrow x \in \pac{1,...,|\subl_1|}$. $x\in \subl_v \Rightarrow x \in \pac{{\sum_{u=1}^{v-1}}|\subl_u|+1,...,\sum_{u=1}^{v}|\subl_u|},\forall 1<v\leq n_{\subl}$. The vertex indexing for the components of the residue is given by: $y\in \supl_{n_\subl+1} \Rightarrow y \in \pac{|V^\subl|+1,...,|V^\subl|+|\supl_{n_\subl+1}|}$. 
$y\in \supl_{v} \Rightarrow y \in \pac{|V^\subl|+ \sum_{u=n_\subl+1}^{v-1}|\supl_u|+1,...,|V^\subl|+ \sum_{u=n_\subl+1}^{v}|\supl_u|}$, $\forall n_\subl+1<v\leq n_\subl+n_\supl$.

The Eqs. \eqref{Eq_SS_TP1} to \eqref{Eq_SS_TPC5} describe frustrating terms with $\bigotimes_{u=1}^{n_\subl}\ket{0}_{\subl_u}$ superstable to the part acting on $V^\subl$ vertices. From these terms, only Eqs. \eqref{Eq_SS_TP2} and \eqref{Eq_SS_TPC1} are of order $\alpha^3\beta$ and thus contribute to the $\beta$ term of the interactions, while the other equations are of order $\alpha^3\beta^2$, contributing to the $\beta^2$ term. In contrast, the terms described by Eqs. \eqref{Eq_SS_TPC9} and \eqref{Eq_Heisenberg_TP} contribute non-superstable terms with $\alpha^3\beta^2$ order. The Eq. \eqref{Eq_SS_TPC9} provides a renormalization of already existing weights in the components on the residue. Instead, Eq. \eqref{Eq_Heisenberg_TP} provides Heisenberg-like connections between the $G_{\rm }\pas{\supl^\prime_u}$ and $G_{\rm }\pas{\supl^\prime_w}$ subgraphs. Therefore, these Heisenberg-like terms add edges and weights that connect different components of $G^{\supl \prime}_\beta$ and that, together with the four cycle contribution, have to lead to the $G^\supl_\beta$ graph by construction.

Similarly, we analyze the $H_{\rm FC}(G)$ terms keeping the indexing. Fourteen possible four cycles emerge from this term and these are encoded in four different cases. To express the relevant terms, we use the notation $A-B-C-D$, which signifies a four cycle that passes through $A$, then $B$, then $C$, then $D$, and then back to $A$.  Here, $A,B,C,D$ can be vertices or components. Furthermore, below, we only provide the factorized expression. The cases are:
\begin{widetext}
\begin{itemize}
    \item {\bf $y-\subl_u-y^\prime-\subl_u$}

\begin{equation}
\begin{split}
 \frac{4}{U^3}\sum_{u,y,y^\prime>y|\pac{y,y^\prime}\notin \subl_u}\pac{t_{\subl_u}(y)t_{\subl_u}(y^\prime)}^2\bigg\{&\frac{|\subl_u|\pap{|\subl_u|+2}}{8} +\frac{|\subl_u|}{2}\pap{6-|\subl_u|}\Vec{S}_y\cdot\Vec{S}_{y^\prime} \\
 &-\bigg[\pap{|\subl_u|-1}\pap{\Vec{S}_y+\Vec{S}_{y^\prime}}+10 i \pap{\Vec{S}_y\times\Vec{S}_{y^\prime}}\bigg]\cdot \Vec{S}_u\\
              &-\frac{\Vec{S}_u\cdot \Vec{S}_u}{2}-10\pap{\Vec{S}_y\cdot\Vec{S}_{y^\prime}}\pap{\Vec{S}_u\cdot \Vec{S}_u}+20 \pap{\Vec{S}_y\cdot\Vec{S}_u} \pap{\Vec{S}_{y^\prime}\cdot \Vec{S}_u}      
              \bigg\}.
\end{split}
\end{equation}
This expression generates the contributions from the four cycles $\subl_u-\subl_v-\subl_u-\subl_v$ (for this case, the term has to be divided over two as it is counted twice), $\subl_u-\subl_v-\subl_u-\subl_w$, $\subl_u-\subl_v-\subl_u-\supl^\prime_w$, $\subl_u-\supl^\prime_v-\subl_u-\supl^\prime_v$, and $\subl_u-\supl^\prime_v-\subl_u-\supl^\prime_w$. The generated terms have order $\alpha^3\beta^2$.

\item {\bf $y-\subl_u-y^\prime-\subl_w$}.
\begin{equation}
\begin{split}
 \frac{4}{U^3}\sum_{u,w>u,y,y^\prime>y|\pac{y,y^\prime}\notin \subl_u\cup\subl_w}&\pac{t_{\subl_u}(y)t_{\subl_u}(y^\prime)t_{\subl_w}(y^\prime)t_{\subl_w}(y)}\bigg\{ \frac{|\subl_u||\subl_w|}{4}-|\subl_u||\subl_w| \Vec{S}_y \cdot \Vec{S}_{y^\prime}\\
 &-\pas{ \pap{\Vec{S}_u|\subl_w|+|\subl_u|\Vec{S}_w}\cdot\pap{\Vec{S}_y+\Vec{S}_{y^\prime}}+\Vec{S}_u\cdot \Vec{S}_w } \\
&+20\pas{ \pap{\Vec{S}_u\cdot\Vec{S}_y} \pap{\Vec{S}_w\cdot \Vec{S}_{y^\prime}}+ \pap{\Vec{S}_w\cdot\Vec{S}_y} \pap{\Vec{S}_u\cdot \Vec{S}_{y^\prime}}-\pap{\Vec{S}_u\cdot \Vec{S}_w}\pap{\Vec{S}_y\cdot\Vec{S}_{y^\prime}} }    \bigg\} .\\
\end{split}
\end{equation}
This expression generates the contributions from four cycles $\subl_u-\subl_v-\subl_w-\subl_x$, $\subl_u-\subl_v-\subl_w-\supl^\prime_x$,$\subl_u-\supl^\prime_v-\subl_w-\supl^\prime_x$, and $\subl_u-\supl^\prime_v-\subl_w-\supl^\prime_v$. The generated terms have order $\alpha^3\beta^2$.

\item{\bf $\subl_u-\subl_u-y-y^\prime$}. Let $\subl_u=\pac{x,x^\prime}$, then the factorized expression reads 
\begin{equation}
\begin{split}
 \frac{2}{U^3}\sum_{u,y>x,y^\prime>x|y^\prime\neq y}\pac{t(x,x^\prime)t_{\subl_u}(y)t(y,y^\prime)t_{\subl_u}(y^\prime)}\bigg\{&\frac{|\subl_u|\pap{|\subl_u|+2}}{4} -|\subl_u|\pap{|\subl_u|+14}\Vec{S}_y\cdot\Vec{S}_{y^\prime} \\
 &-\Vec{S}_u\cdot \Vec{S}_u-2\pap{|\subl_u|-1}\pap{\Vec{S}_y+\Vec{S}_{y^\prime}}\cdot \Vec{S}_u\\
              &+20\pap{\Vec{S}_y\cdot\Vec{S}_{y^\prime}}\pap{\Vec{S}_u\cdot \Vec{S}_u}      
              \bigg\}.
\end{split}
\end{equation}
This expression generates the contributions from the four cycles $\subl_u-\subl_u-\subl_w-\subl_w$, $\subl_u-\subl_u-\subl_w-\supl^\prime_x$ (the term generated by this cycle has to be multiplied by two), and $\subl_u-\subl_u-\supl^\prime_w-\supl^\prime_w$. The generated terms have order $\alpha^3\beta$ or $\alpha^3\beta^{3/2}$. The latter order is included in our demonstration by simply defining a new parameter $\beta^\prime:=\sqrt{\beta}$ to calculate the sufficient condition that leads to a superstable ground state. 

\item {\bf $\subl_u-\subl_w-v-v^\prime$}

\begin{equation}
\begin{split}
 \frac{2}{U^3}\sum_{u,w\neq u,y>\sum_{v=1}^{u}|\subl_v|,y^\prime>\sum_{v=1}^{w}|\subl_v||y^\prime\neq y}&\pac{t_{u-w}t_w(y^\prime)t(y^\prime,y)t_u(y)}\bigg\{\frac{|\subl_u||\subl_w|}{4}-|\subl_u||\subl_w| \Vec{S}_y \cdot \Vec{S}_{y^\prime}\\
 & -\pas{ \pap{\Vec{S}_u|\subl_w|+|\subl_u|\Vec{S}_w}\cdot\pap{\Vec{S}_y+\Vec{S}_{y^\prime}}+\Vec{S}_u\cdot \Vec{S}_w } \\
&+20\pas{\pap{\Vec{S}_u\cdot \Vec{S}_w}\pap{\Vec{S}_y\cdot\Vec{S}_{y^\prime}}+ \pap{\Vec{S}_u\cdot\Vec{S}_y} \pap{\Vec{S}_w\cdot \Vec{S}_{y^\prime}}- \pap{\Vec{S}_w\cdot\Vec{S}_y} \pap{\Vec{S}_u\cdot \Vec{S}_{y^\prime}} }    \bigg\} .
\end{split}
\end{equation}
This expression generates the contributions from the four cycles $\subl_u-\subl_u-\subl_w-\subl_x$, and $\supl^\prime_u-\supl^\prime_u-\subl_w-\subl_x$. The generated terms have order $\alpha^3\beta^{3/2}$ or $\alpha^3\beta^2$. 
\end{itemize}
\end{widetext}
The first line of the corresponding generating equation renormalizes already existing weights for the four cycles $\subl_u-\supl^\prime_v-\subl_u-\supl^\prime_v$, $\subl_u-\supl^\prime_v-\subl_w-\supl^\prime_v$, $\subl_u-\subl_u-\supl^\prime_w-\supl^\prime_w$, and $\supl^\prime_u-\supl^\prime_u-\subl_w-\subl_x$. Meanwhile, the first line of the corresponding generating equation for the four cycles $\subl_u-\supl^\prime_v-\subl_u-\supl^\prime_w$, and $\subl_u-\supl^\prime_v-\subl_w-\supl^\prime_x$ provides weights between vertices from different components of $G^{\supl \prime}_\beta$. These terms create Heisenberg-like $\beta$-dependent couplings that are not part of the original graph. All the other four cycles and lines of the generating equations contribute terms that are part of the frustrating term, as they all contain an $S_{u}^\alpha$ operator.
 
Hence, we have shown that features identified by our algorithm lead to superstable graphs identified by the first and third  SW transformation expansion order, as one recovers an effective Hamiltonian with the form of a $H^\sub(\beta)$. In particular, we have shown that $\bigotimes_{u=1}^{n_\subl}\ket{0}_{\subl_u}$ is superstable to the frustrating term and that new edges are added by higher-order expansion terms.

 \section{Adiabatic continuation to different superstable graphs for the diamond chain}
\label{APP_AdiabaticConection}
Here we show the existence of adiabatic continuation between the ground state of the regimes $\mathrm{II}$ and $\mathrm{III}$ of the diamond chain and the superstable ground state of different superstable graphs. In particular, we seek an adiabatic continuation that preserves the total spin, allowing us to apply Theorem 1 to the connected superstable graph and determine the magnetization in the ground state regimes $\mathrm{II}$ and $\mathrm{III}$. Such an adiabatic continuation is possible when the ground states of the initial and final associated spin Hamiltonians are unique, and we can interpolate between the two while preserving the total spin and uniqueness of the ground state (up to the mandatory degeneracy). 


The conditions for adiabatic continuation are fulfilled in the $n=1$  SW transformation expansion order. Let $G_{\rm T}\pap{t_\subl}$ be the graph describing the diamond chain in the MT. We define $J(G)=(V,E,4t(e)^2/U)$ as the map connecting the $G_{\rm H}$ graph with the graph $G$. At this expansion order, the total spin of each core, $S_u$, is a good quantum number of the effective spin Hamiltonian. That is, all eigenstates $\ket{j}$ of  $\Hham (G_{\rm H})$ give $\Vec{S}_u \cdot \Vec{S}_u \ket{j}=S_u(S_u+1)\ket{j}\forall u \in \mathcal{I_\subl}$. We next make the following assumptions: (i) the ground state of $\Hham\pap{J\pap{G_{\rm T}\pap{t_\subl}}}$, $\ket{0}$, is unique with gap $\Delta>0$ in the smallest $|m_z|$ (the quantum number for the spin projection along the $z$-axis) sector away from the regime boundaries. (ii) the $S_u$ values of $\ket{0}$ are known for all $u$. In Lieb's regime, these conditions are given by the Lieb-Mattis (uniqueness) and Shen-Qiu-Tian ($S_u=1\forall u$) theorems. In the tetramer-dimer regime, the use of these assumptions is justified by numerical \cite{Long_1990_TetramerDimerOpenNumerical} and analytical findings with periodic boundary conditions \cite{Takano_Triangular_Clusters,Triangular_Heisenberg}. Within this regime, the $S_u$ quantum numbers alternate between $0$ and $1$ along the diamond chain. 

Consider the Hamiltonian
\begin{equation}
\begin{split}  
\label{Eq_APPAdiabaticTriangular}
H_{\rm A}\pap{\alpha,E_1,t_\subl}:=&\Hham\pap{J\pap{G_{\rm T}\pap{t_\subl}}}\\
&-\frac{4\alpha t_\subl^2}{U}\sum_{(x,x^\prime)\in E_1} \pap{\Vec{S}_x \cdot \Vec{S}_{x^\prime} -\frac{1}{4}},
\end{split}
\end{equation}
where $\alpha\in\pas{0,1}$. $E_1\subseteq E^\subl$ is the subset of edges coupling the vertices of the dimers of $G^\subl$ that contribute $S_u=1$ to the ground state $\ket{0}$ at $t_\subl$. Here, $E_1$ is composed of all the vertical edges in Lieb's regime and every second edge in the tetramer-dimer regime. Thus, $H_{\rm A}\pap{1,E_1,t_\subl}$ is $\Hham\pap{J\pap{G_{\rm T}\pap{t_\subl}}}$ without the edges connecting the dimers with $S_u=1$. A graph in the Hubbard model leading to a $n=1$ effective spin Hamiltonian given by $H_{\rm A}\pap{1,E_1,t_\subl}$ is obtained by removing the corresponding vertical edges from the diamond chain. In both regimes, such a graph is superstable; removing all vertical edges corresponds to a bipartite graph and removing every second vertical edge  generates balanced components of $G\pas{V^\supl}$. Therefore, $H_{\rm A}\pap{1,E_1,t_\subl}$ has the form of a $H^\sub(\beta)$ Hamiltonian.

Since $S_u$ is a quantum number of $\Hham\pap{J\pap{G_{\rm T}\pap{t_\subl}}}$ and each term in the sum in Eq. \eqref{Eq_APPAdiabaticTriangular} is proportional to $S_u(S_u+1)$, then all eigenstates $\ket{j}$ of $\Hham\pap{J\pap{G_{\rm T}\pap{t_\subl}}}$ are eigenstates of $H_{\rm A}\pap{\alpha,E_1,t_\subl}$. The ground state of $\Hham\pap{J\pap{G_{\rm T}\pap{t_\subl}}}$ ($\ket{0}$)  yields to the expected value
\begin{equation}
\begin{split}
   \bra{0}{&H_{\rm A}\pap{\alpha,E_1,t_\subl}}\ket{0}=\min\pap{\Hham\pap{J\pap{G_{\rm T}\pap{t_\subl}}}}\\
   &-\frac{4\alpha t_\subl^2}{U}\sum_{(x,x^\prime)\in E_1}\max\pap{\Vec{S}_x \cdot \Vec{S}_{x^\prime} -\frac{1}{4}}, 
\end{split}
\end{equation}
which saturates the variational principle's inequality. Thus, $\ket{0}$ is also a ground state of $H_{\rm A}\pap{\alpha,E_1,t_\subl}$. Now, consider any $\ket{j\neq0}$. These lead to
\begin{equation}
\begin{split}
   \bra{j\neq0}{H_{\rm A}\pap{\alpha,E_1,t_\subl}}&\ket{j\neq0}\geq\min\pap{\Hham\pap{J\pap{G_{\rm T}\pap{t_\subl}}}}+\Delta\\
   &-\frac{4\alpha t_\subl^2}{U}\sum_{(x,x^\prime)\in E_1}\max\pap{\Vec{S}_x \cdot \Vec{S}_{x^\prime} -\frac{1}{4}}, 
\end{split}
\end{equation}
which is strictly larger than $\bra{0}{H_{\rm A}\pap{\alpha,E_1,t_\subl}}\ket{0}$.

Thus, at the smallest $|m_z|$ sector and away from the regime boundaries, $\ket{0}$ is the only ground state of $H_{\rm A}\pap{\alpha,E_1,t_\subl}\forall \alpha$. Since $\ket{0}$ eliminates the couplings frustrating the cores of the graph obtained at $\alpha=1$, it is the superstable ground state of $H_{\rm A}\pap{1,E_1,t_\subl}$. Therefore, $\ket{0}$ can be continued to the superstable ground state of a $H^\sub(\beta)$ given by $H_{\rm A}\pap{1,E_1,t_\subl}$ through the path defined by $H_{\rm A}\pap{\alpha,E_1,t_\subl}$. Moreover, since $H_{\rm A}\pap{1,E_1,t_\subl}$ is the effective spin model obtained from a superstable graph, we can determine the magnetization of the Lieb and the tetramer-dimer regimes of the diamond chain through adiabatic continuation to the superstable ground state of two superstable graphs; one is given by the graph $G_{\rm T}\pap{0}$ and the other by removing every second vertical edge of the diamond chain.
\newpage
\bibliography{Doped_Hubbard}
\end{document}